\documentclass[8.5pt,twoside,twocolumn]{article}

\usepackage[super,sort&compress,comma]{natbib} 
\usepackage{mhchem}
\usepackage{times,mathptmx}
\usepackage{amsmath,amsfonts}
\usepackage{amssymb}
\usepackage{sectsty}
\usepackage{balance} 
\usepackage[usenames,dvipsnames]{xcolor}
\usepackage{graphicx} 
\usepackage{lastpage}
\usepackage[format=plain,justification=raggedright,singlelinecheck=false,font=small,labelfont=bf,labelsep=space]{caption} 
\usepackage{fancyhdr}
\usepackage[latin1]{inputenc}
\begin{document}

\thispagestyle{plain}
\fancypagestyle{plain}{
\renewcommand{\headrulewidth}{1pt}}
\renewcommand{\thefootnote}{\fnsymbol{footnote}}
\renewcommand\footnoterule{\vspace*{1pt}%
\hrule width 3.4in height 0.4pt \vspace*{5pt}} 
\setcounter{secnumdepth}{5}

\newcommand{\blue}[1]{\textcolor{blue}{#1}}
\newcommand{\green}[1]{\textcolor{green}{#1}}
\newcommand{\olivegreen}[1]{\textcolor{OliveGreen}{#1}}
\newcommand{\red}[1]{\textcolor{red}{#1}}
\definecolor{violet}{rgb}{0.55,0.14,1.0}
\newcommand{\violet}[1]{\textcolor{violet}{#1}}
\newcommand{\purple}[1]{\textcolor{Purple}{#1}}
\newcommand{\cyan}[1]{\textcolor{cyan}{#1}}
\newcommand{\magenta}[1]{\textcolor{magenta}{#1}}

\newcommand{\Q}[1]{\textcolor{red}{#1}}
\newcommand{\s}[1]{\textcolor{green}{#1}}
\newcommand{\sw}[1]{\textcolor{green}{\small{#1}}}

\makeatletter 
\def\subsubsection{\@startsection{subsubsection}{3}{10pt}{-1.25ex plus -1ex minus -.1ex}{0ex plus 0ex}{\normalsize\bf}} 
\def\paragraph{\@startsection{paragraph}{4}{10pt}{-1.25ex plus -1ex minus -.1ex}{0ex plus 0ex}{\normalsize\textit}} 
\renewcommand\@biblabel[1]{#1}            
\renewcommand\@makefntext[1]%
{\noindent\makebox[0pt][r]{\@thefnmark\,}#1}
\makeatother 
\renewcommand{\figurename}{\small{Fig.}~}
\sectionfont{\large}
\subsectionfont{\normalsize} 

\fancyfoot{}
\fancyfoot[RO]{\footnotesize{\sffamily{1--\pageref{LastPage} ~\textbar  \hspace{2pt}\thepage}}}
\fancyfoot[LE]{\footnotesize{\sffamily{\thepage~\textbar\hspace{3.45cm} 1--\pageref{LastPage}}}}
\fancyhead{}
\renewcommand{\headrulewidth}{1pt} 
\renewcommand{\footrulewidth}{1pt}
\setlength{\arrayrulewidth}{1pt}
\setlength{\columnsep}{6.5mm}
\setlength\bibsep{1pt}

\twocolumn [
  \begin{@twocolumnfalse}
\noindent\LARGE{\textbf{Yielding of binary colloidal glasses}}
\vspace{0.6cm}

\noindent\large{\textbf{T. Sentjabrskaja,\textit{$^{a}$} E. Babaliari,$\textit{$^{b}$}$ J. Hendricks,\textit{$^{a}$} M. Laurati,$^{\ast}$\textit{$^{a}$} G. Petekidis$\textit{$^{b}$}$  and
S.U. Egelhaaf~\textit{$^{a}$}}}\vspace{0.5cm}

\noindent\textit{\small{\textbf{Received Xth XXXXXXXXXX 20XX, Accepted Xth XXXXXXXXX 20XX\newline
First published on the web Xth XXXXXXXXXX 200X}}}

\noindent \textbf{\small{DOI: 10.1039/b000000x}}
\vspace{0.6cm}

\noindent \normalsize{The rheological response, in particular the non-linear response, to oscillatory shear is experimentally investigated in colloidal glasses. The glasses are highly concentrated binary hard-sphere mixtures with relatively large size disparities. For a size ratio of 0.2, a strong reduction of the normalized elastic moduli, the yield strain and stress and, for some samples, even melting of the glass to a fluid is observed upon addition of the second species. This is attributed to the more efficient packing, as indicated by the shift of random close packing to larger total volume fractions. This leads to an increase in free volume which favours cage deformations and hence a loosening of the cage. Cage deformations are also favoured by the structural heterogeneity introduced by the second species. For a limited parameter range, we furthermore found indications of two-step yielding, as has been reported previously for attractive glasses. In samples containing spheres with more comparable sizes, namely a size ratio of 0.38, the cage seems less distorted and structural heterogeneities on larger length scales seem to become important. The limited structural changes are reflected in only a small reduction of the moduli, yield strain and stress.}
\vspace{0.5cm}
 \end{@twocolumnfalse}
  ]



\footnotetext{\textit{$^{a}$~Condensed Matter Physics Laboratory, Heinrich-Heine University, D\"usseldorf, Germany. E-mail: marco.laurati@uni-duesseldorf.de}}
\footnotetext{\textit{$^{b}$~FORTH/IESL and Department of Materials Science and Technology, University of Crete, 71110, Heraklion, Greece.}}



\section{Introduction}

Many particle dispersions used in applications, for example paint, ink, cement, ceramics or foodstuffs, are characterised by a size distribution of the dispersed phase. 
Even if a monodisperse system is desirable, it is often difficult to avoid a distribution of particle sizes. 
Furthermore, through the size distribution, the properties of a dispersion, such as its rheological behaviour, can be tuned, for instance to meet processing or application needs. 
To investigate the effect of a distribution of sizes, binary mixtures of spherical colloidal particles represent the simplest model system.

The interactions and the phase behaviour of binary colloidal hard-sphere mixtures have been studied by theory\cite{amokrane:2005,roth:2000,ashton:2011,dijkstra:99,goetzelmann:99} and simulations.\cite{dijkstra:99,goetzelmann:99,malherbe:2007}
In equilibrium, binary colloidal mixtures exhibit a wider fluid-solid coexistence region than one-component systems, which has been thoroughly investigated in experiments.\cite{van_duijnevelt:93,dinsmore:95,imhof:prl:95,hunt:2000} Additionally, formation of complex crystalline structures through co-crystallisation of the two species is predicted and observed.\cite{bartlett:92,cottin:95,schofield:2001,hynninen:09} 
For size ratios $\delta = R_S/R_L \lesssim 0.2$, where $R_S$ and $R_L$ are the radii of the small and large spheres, respectively, theory expects fluid-fluid and solid-solid coexistences,\cite{dijkstra:99} which are also observed in simulations\cite{dijkstra:98} but not yet in experiments. 
In addition, non-equilibrium glass states have been predicted theoretically\cite{voigtmann:2003,voigtmann:epl:2011,germain:2009,amokrane:2012} and observed experimentally.\cite{imhof:prl:95,williams:2001}
In particular, Mode Coupling Theory (MCT) predicts that, at constant total volume fraction $\phi$, a one-component glass is melted upon addition of a sufficient amount of spheres with a different size ($\delta\le 0.65$).\cite{voigtmann:2003,voigtmann:epl:2011}
This is consistent with the faster structural relaxation experimentally observed in samples with $\delta\approx0.6$, $\phi\approx0.58$ and intermediate mixing ratios.\cite{williams:2001}
This leads to a strong decrease of the viscosity, which has been determined in experiments and simulations for a sufficiently large degree of mixing.\cite{rodriguez:92, foffi:2003}
Recent MCT results\cite{voigtmann:epl:2011} furthermore predict that for a large size disparity, $\delta\leq0.2$, different glass states exist, which are distinguished by caging of one or both species, or by depletion induced bonding of the large spheres. 
The latter, for which some experimental evidence exists for $\delta\approx 0.1$,\cite{imhof:prl:95} is expected to show similarities with attractive glasses as those observed in colloid-polymer mixtures.\cite{pham02,pham04}

Similar to the interactions and the phase behavior, also the rheological response of binary mixtures changes upon varying the size and mixing ratio\s{s}.
This has been studied experimentally,\cite{woutersen:93,mewis:94,richtering:95,zukoski:96,gondret:97,greenwood:97,shikata:98,rodriguez:92} theoretically\cite{ohtsuki:83,nagele:98,lionberger:02} and by simulations.\cite{chang:93,chang:94,powell:94}
In the granular limit, i.e.~when Brownian motion becomes irrelevant, binary mixtures with a size ratio $\delta = 0.2$ exhibit a minimum of the viscosity at a relative volume fraction of small spheres, $x_{\mathrm{S}} \approx 0.4$,\cite{farris:68} which is known as \textit{Farris} effect. 
In contrast, for colloidal mixtures a minimum of the viscosity is only observed at high total volume fractions $\phi\ge0.4$ and at a mixing ratio which depends on $\phi$ and $\delta$.\cite{lionberger:02} 
With decreasing $\delta$, the minimum occurs at smaller fractions of small spheres, which results from a balance between the more efficient packing, since small spheres can fill the space between large spheres, and the depletion attraction induced between large spheres.\cite{lionberger:02}
Nevertheless, the rheology of concentrated binary colloidal mixtures has hardly been studied,\cite{rodriguez:92,foudazi:2012} especially of spheres with significantly different sizes, i.e.~small size ratios $\delta$.

Here we investigate the rheology of dispersions containing binary mixtures with small size ratios, $\delta\approx0.2$ and 0.38, over a broad range of total volume fractions $\phi$ and mixing ratios, characterized by the relative volume fraction of small spheres $x_{\mathrm{S}} = \phi_\mathrm{S}/\phi$.
Their response to oscillatory shear is studied with a particular focus on the non-linear viscoelastic properties, while the linear response, together with the structure and dynamics at rest, will be discussed in detail elsewhere.\cite{sentjabrskaya_linear_2012,sentjabrskaya_linear_2012_2}
In the present case of spheres with significantly different sizes (i.e.~small $\delta$), the non-linear response contains contributions related to the different length scales present in the samples. 
This is similar to colloid-polymer mixtures, where systems with attractive interactions, such as gels or attractive glasses, are characterized by two yielding processes.\cite{pham06,pham08,laurati:jor:2011,koumakis:soft:2011}
The two yielding processes reflect the breaking of inter-particle `bonds' and cluster breaking, in the case of gels, or irreversible cage deformation, in the case of attractive glasses.\cite{laurati:jor:2011,koumakis:soft:2011}
The yielding behaviour of attractive systems is hence different from the one of repulsive systems, which typically only show one yielding mechanism related to cage distortion.\cite{petekidis02,petekidis03,pham06,pham08}

\section{Materials and Methods}

\subsection{\label{rheo_meas}Rheology}

Rheological measurements were performed with an AR2000ex stress-controlled rheometer, and ARES G2 and ARES strain-controlled rheometers from TA instruments, using cone and plate geometries of diameter $D=20$~mm, cone angle $\alpha=2^{\circ}$ and gap $d=0.054$~mm (AR2000ex), $D=25$~mm, $\alpha=2^{\circ}$ and $d=0.048$~mm (ARES G2) and $D=25$~mm and 50~mm, $\alpha=2^{\circ}$ and $d=0.048$~mm (ARES). 
Solvent traps were used in all rheometers to minimize solvent evaporation. 
The temperature was set to $T=20\;^{\circ}$C and controlled within $\pm 0.1\;^{\circ}$C via a standard Peltier plate (AR2000ex, ARES) or an advanced Peltier system (ARES G2). 
The effects of sample loading and aging were reduced by performing the following rejuvenation procedure before each test.
Directly after loading, a dynamic strain sweep was performed to estimate the strain amplitude $\gamma$ at which the system starts to flow, i.e.~oscillatory shear was applied to the samples with frequency $\omega = 1$~rad/s and increasing $\gamma$ until the sample showed a liquid-like response.  
Then, before each measurement, flow of the sample was induced by applying oscillatory shear at a sufficiently large strain. 
In the case of the size ratio $\delta = 0.20$, $\gamma = 300$\% was used for all samples. 
For $\delta = 0.38$, different values $200$\%~$\le \gamma \le 1000$\% were used depending on the volume fraction $\phi$ and relative volume fraction of small particles $x_{\mathrm{s}}$. 
Shear was applied until a steady-state response, i.e.~a time-independent storage $G^\prime$ and loss modulus $G^{\prime\prime}$, was achieved, which typically took about 200~s. 
Subsequently the samples were sheared at $0.1$\%~$\leq\gamma\leq 1.5$\% (depending on sample) until the linear viscoelastic moduli reached a time-independent value, typically after 100~s to 900~s (depending on sample). 
This indicated that no further structural changes occurred and hence a reproducible state of the sample was reached and a new measurement could be started. Note that ageing effects might be present at longer waiting times.
Measurements with serrated and smooth geometries, respectively, yielded comparable results suggesting the absence of wall slip.

\subsection{\label{samples}Samples}

\begin{figure}[h]
\centering
   \includegraphics[scale=0.25]{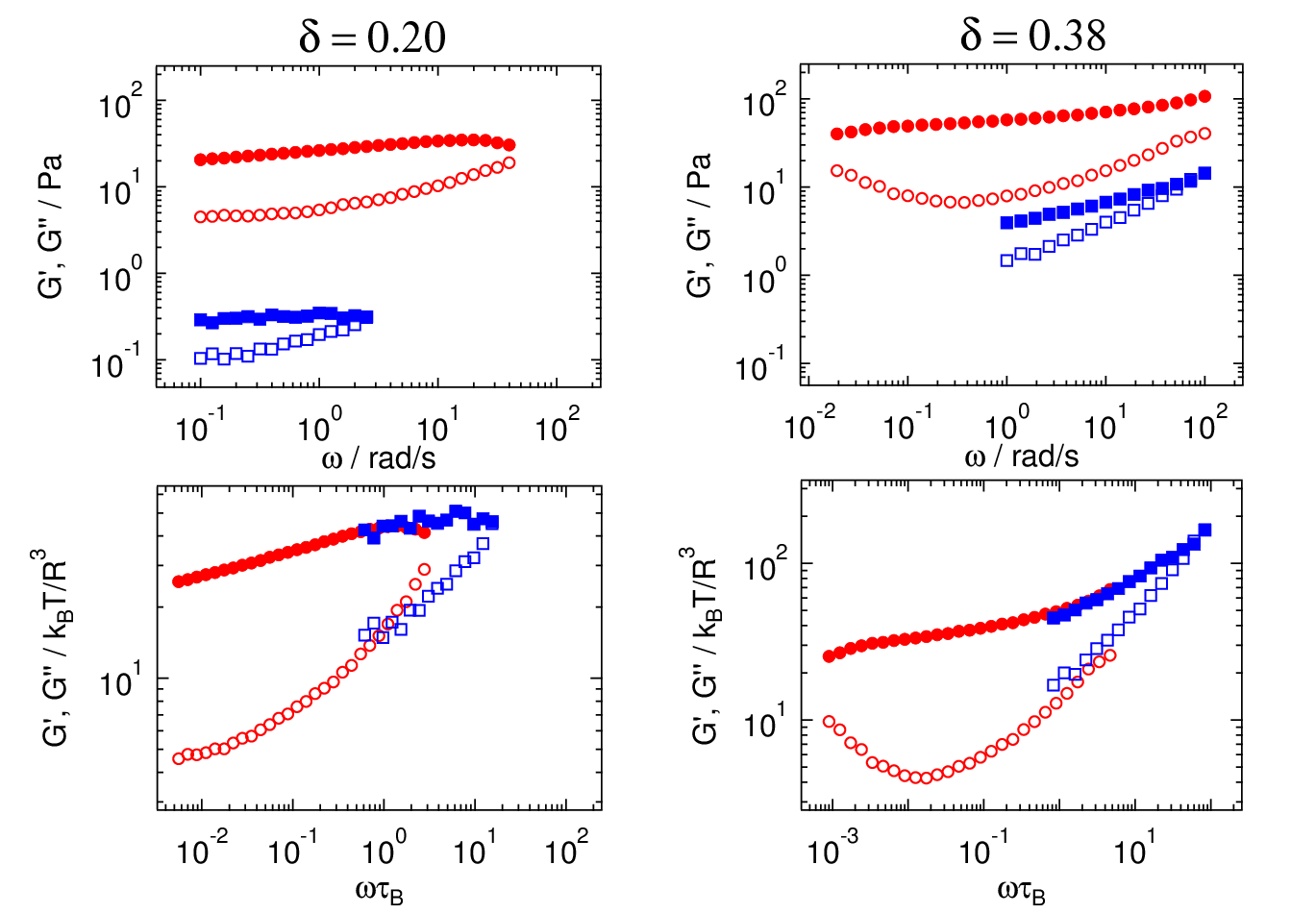}
  \caption{(top) Storage, $G^{\prime}$ (full symbols), and loss, $G^{\prime\prime}$ (open symbols), moduli of samples containing large (\blue{$\Box$}) and small (\red{\Large{$\circ$}}) spheres, respectively, as a function of frequency $\omega$ obtained by Dynamic Frequency Sweep measurements for (left) a size ratio $\delta= 0.20$ and total volume fraction $\phi = 0.58$ and (right) $\delta=0.38$ and $\phi = 0.595$. (bottom) Same data in units proportional to the energy density, i.e.~$k_{\mathrm{B}}T/R^3$, and Brownian time $\tau_B = D_0/R^2$. The strain amplitude was $\gamma = 0.5$\% for $\delta= 0.20$ and $\gamma = 1.5$\% for $\delta=0.38$.}
  \label{fig1:phi_adjust_rheo}
\end{figure}

Polymethylmethacrylate (PMMA) spheres sterically stabilized with a layer of polyhydroxystearic acid (PHSA) were dispersed in a mixture of cycloheptyl bromide (CHB) and cis-decalin that closely matched the density and refractive index of the colloids ($\delta = 0.20$ and $0.19$) or in a mixture of octadecene and bromonaphtalene which minimizes solvent evaporation ($\delta = 0.38$). 
For samples in octadecene-bromonaphtalene, measurements of the time evolution of the linear viscoelastic moduli indicate  the absence of significant gravitational effects over times much longer than typical measurement times.
In the CHB/decalin mixture, the particles acquire a small charge which was screened by adding 4~mM tetrabutylammoniumchloride.\cite{yethiraj03} 
In this case, the colloids behave hard-sphere-like in both solvent mixtures.
PMMA spheres with different average radii were used; $R_{\mathrm{L}}^{\mathrm{F}}$ = 880~nm (polydispersity 0.057) and $R_{\mathrm{S1}}$ = 175~nm (polydispersity 0.150) to result in $\delta = 0.20$,  $R_{\mathrm{L}}^{\mathrm{NF}}$ = 942~nm (polydispersity 0.06) and the same $R_{\mathrm{S1}}$ to result in $\delta = 0.19$, $R_{\mathrm{L}}$ = 358~nm (polydispersity 0.140) and $R_{\mathrm{S2}}$ = 137~nm (polydispersity 0.120) to result in $\delta = 0.38$.
The radii and polydispersities were determined from the angular dependencies of the scattered intensity and the diffusion coefficients, obtained using static and dynamic light scattering, respectively, with very dilute colloidal suspensions ($\phi\simeq 10^{-4}$). 
For the large spheres, a similar radius, $R_{\mathrm{L}}^{\mathrm{F}} = 885$~nm, has been estimated from the position of the first peak of the radial distribution function, which was obtained by confocal microscopy.\cite{jenkins08}
Confocal microscopy could be performed with these large spheres, because they were fluorescently labelled with nitrobenzoxadiazole (NBD). 
Confocal microscopy was also used to determine the volume fraction of a dispersion of these spheres as follows.
A random close packed sample was obtained by sedimenting a dilute suspension in a centrifuge.
The sediment, whose volume fraction was roughly estimated using simulation results,\cite{schaertl94} was subsequently diluted to a volume fraction $\phi\simeq 0.4$ and imaged by confocal microscopy. 
The imaged volume was partitioned into Voron{\"o}i cells and their mean volume determined. 
The ratio of the particle volume to the mean Voron{\"o}i volume provides an estimate of the volume fraction of the sample, $\phi = 0.43$.
This allowed us to calculate the volume fraction of the random close packed stock solution $\phi_{\mathrm{RCP}}^{\mathrm{L}}= 0.68$. 
The smaller spheres were too small to be imaged (thus also not fluorescently labelled).
The volume fraction of their sediment was estimated taking into account their polydispersity:\cite{schaertl94} $\phi_{\mathrm{RCP}}^{\mathrm{S1}}\simeq 0.68$ for spheres with radius $R_{\mathrm{S1}}=175$~nm and $\phi_{\mathrm{RCP}}^{\mathrm{S2}}\simeq \phi_{\mathrm{RCP}}^{\mathrm{L}}\simeq 0.67$ for spheres with radii $R_{\mathrm{S2}}=137$~nm and $R_{\mathrm{L}}=358$~nm.
The value of the volume fraction is known to suffer from relatively large uncertainties.\cite{Poon/Weeks/Royall} 
Thus the value of $\phi$ obtained for the large spheres was used as a reference value and the volume fraction of the two batches containing the smaller particles adjusted using rheological measurements as follows. 
Linear viscoelastic moduli for samples at nominally equal volume fraction ($\phi \simeq 0.58$ for $\delta = 0.20$, $\phi \simeq 0.595$ for $\delta = 0.38$, $\phi \simeq 0.61$ for $\delta = 0.19$) were measured in Dynamic Frequency Sweeps (DFS) at a strain amplitude $0.1$\%~$\leq\gamma\leq 1.5$\% (depending on sample). 
The obtained storage moduli $G^{\prime}$ and loss moduli $G^{\prime\prime}$ as a function of oscillation frequency $\omega$ are expected to agree for spheres of different size but the same volume fraction, if the moduli are rescaled by the energy density $\sim k_{\mathrm{B}}T/R^3$ and the frequency by the Brownian time $\tau_{\mathrm{B}}=R^2/D_0$ with $D_0 = 6\pi\eta R$ the Stokes-Einstein-Sutherland diffusion coefficient in the dilute limit\cite{koumakis:2012} and $\eta$ the solvent viscosity.
The dispersion of small spheres was diluted until its rescaled linear response matched that of the dispersion of large spheres with the desired volume fraction, i.e.~until an equivalent rheological response in the linear regime was obtained (Fig.~\ref{fig1:phi_adjust_rheo}).
Furthermore, it was verified that the normalised elastic modulus $G^{\prime}$ and its $\phi$ dependence coincides, for all particles used, with that of a dispersion containing crystallising colloids with a low polydispersity, whose volume fraction was determined in the crystal-fluid coexistence region.\cite{koumakis:2012}
If illuminated by laser light, Bragg reflections were not observed, indicating the absence of crystallinity in the one-component dispersions.
By mixing appropriate amounts of the one-component dispersions, samples with different total volume fractions $\phi$ and relative volume fractions of small particles $x_{\mathrm{s}} = \phi_{\mathrm{S}}/(\phi_{\mathrm{S}}+\phi_{\mathrm{L}})$ were prepared, where $\phi_{\mathrm{S}}$ and $\phi_{\mathrm{L}}$ are the volume fractions of small and large particles, respectively.
Samples with different $x_{\mathrm{S}}$ and two different values of $\phi$ (for $\delta = 0.20$, 0.38) as well as fixed $x_{\mathrm{S}} = 0.65$ and different values of $\phi$ (for $\delta = 0.19$) were investigated.
%


\section{Results and Discussion}

\begin{figure*}[!ht]
\centering
   \includegraphics[scale=0.6]{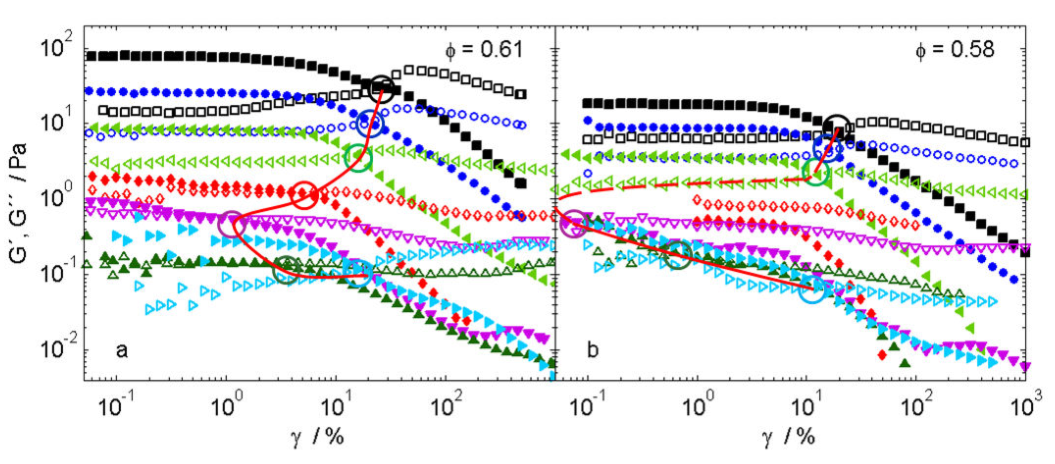}
  \caption{(a,b) Storage, $G^{\prime}$, (full symbols) and loss, $G^{\prime\prime}$, (open symbols) moduli as a function of the strain amplitude $\gamma$ obtained in DSS measurements. The size ratio $\delta = 0.20$, the total volume fraction (a) $\phi=0.61$ and (b) $\phi=0.58$, the relative volume fraction of small particles $x_{\mathrm{S}}= 0.0$ (\cyan{$\triangleright$}), 0.1 (\olivegreen{$\triangle$}), 0.3 (\purple{$\bigtriangledown$}), 0.5 (\red{$\Diamond$}), 0.7 (\green{$\triangleleft$}), 0.9 (\blue{{\Large ${\circ}$}}), 1.0 ($\Box$) and $Pe_{\omega}=5.55\times 10^{-1}$ (corresponding to $7.6\times 10^{-2}$~rad/s~$\leq \omega \leq 9.7$~rad/s). Circles indicate the yields strains $\gamma_{\mathrm{y}}$ and the red solid line their $x_{\mathrm{S}}$-dependence.}
  \label{fig2:dss_diff_xs}
\end{figure*}

In Dynamic Strain Sweep (DSS) experiments, a sinusoidal strain is applied whose frequency $\omega$ is constant but whose amplitude $\gamma$ is increased in steps, starting in the linear viscoelastic regime and progressing into the non-linear regime. 
The stress response of the system is recorded as a function of strain amplitude $\gamma$.
Figure \ref{fig2:dss_diff_xs} shows results of DSS measurements for samples with size ratio $\delta = 0.20$, total volume fractions $\phi = 0.61$ and $0.58$, and different relative volume fractions of small spheres $x_{\mathrm{S}}$. 
Beyond the linear viscoelastic regime the stress response in DSS experiments significantly deviates from a simple sinusoidal form and can be decomposed into higher order (odd) harmonics, as shown before for one-component hard-sphere glasses.\cite{koumakis:2012}.
However, the $G^{\prime}$ and $G^{\prime\prime}$ values shown in figure \ref{fig2:dss_diff_xs} correspond to the first harmonic contribution of the stress response.
To allow for a comparison of the different samples, measurements were not performed at a constant frequency $\omega$, but at a fixed oscillatory Peclet number $Pe_{\omega}=\omega\tau_{\mathrm{B}}$.
It is the ratio of the Brownian time of the system, $\tau_B=\langle R^2/D_0\rangle$, and the timescale imposed by shear, i.e.~the inverse of the frequency, $\tau_{\omega}=1/\omega$.
Thus, $Pe_{\omega} = \langle (6\pi\eta R^3)/(k_{\mathrm{B}}T)\rangle \omega$ and
\begin{equation}
\left\langle R^3 \right\rangle = R_{\mathrm{L}}^3\left[x_{\mathrm{S}}\left(\frac{1}{\delta^3}-1\right)+1\right]^{-1}   \;\; .
\label{eq:R3}
\end{equation}
We applied $Pe_{\omega} = 5.55\times 10^{-1}$ corresponding to $7.6\times 10^{-2}$~rad/s~$\leq \omega \leq 9.7$~rad/s, depending on $x_{\mathrm{S}}$.

The one component systems ($x_S=0$ and 1) for both $\phi$ show the characteristic response of a hard sphere glass (Fig.~\ref{fig2:dss_diff_xs}a,b).\cite{petekidis03,pham06,pham08,koumakis:2012}
(Note that due to the much lower energy density of the samples with the large spheres, their response is much weaker and thus more affected by noise.)
The storage modulus $G^{\prime}$ is larger than the loss modulus G$^{\prime\prime}$ in the linear viscoelastic regime, with their values comparable to the ones obtained in dynamic frequency sweeps (Fig.~\ref{fig1:phi_adjust_rheo}).\cite{sentjabrskaya_linear_2012}
The two moduli become equal at a strain amplitude $\gamma_{\mathrm{y}}$ (highlighted with circles in Fig.~\ref{fig2:dss_diff_xs}), which is identified with the yield strain of the glass.
At the yield strain $\gamma_{\mathrm{y}}$ and the corresponding yield stress $\sigma_{\mathrm{y}}$, the local environment of a particle is irreversibly rearranged, i.e.~its cage broken.\cite{petekidis02,petekidis03,pham06,pham08} 
For $\gamma > \gamma_{\mathrm{y}}$, $G^{\prime\prime}$ is larger than $G^{\prime}$ and the system starts to flow.
In this regime, $G^{\prime\prime}$ shows a maximum which indicates the largest energy dissipation and has previously also been used to estimate the yield strain associated with irreversible rearrangements of the cage.\cite{petekidis02,pham08}
Upon increasing the volume fraction from $\phi=0.58$ to 0.61, the linear viscoelastic moduli and the yield strain $\gamma_{\mathrm{y}}$ increase.
This is consistent with previous studies,\cite{petekidis02,petekidis03,koumakis:2012} which found $\gamma_{\mathrm{y}}$ to increase with volume fraction up to $\phi \approx 0.62$, beyond which it decreases due to the approach toward random close packing.


Keeping the total volume fraction $\phi$ constant, but changing the composition to $x_{\mathrm{S}}=0.9$, the storage and loss moduli decrease (Fig.~\ref{fig2:dss_diff_xs}). 
The decrease is not only due to the presence of large particles and hence a lower energy density, but remains even if the moduli are rescaled by the energy density $\langle n\,k_{\mathrm{B}}T \rangle \sim  1/\langle R^3 \rangle$.
This indicates a softening of the glass.
A softer response is also reflected in a reduced yield strain $\gamma_{\mathrm{y}}$ and yield stress $\sigma_{\mathrm{y}}$ (circles in Fig.~\ref{fig2:dss_diff_xs}.
A further decrease of the relative volume fraction of small spheres to $x_{\mathrm{S}}=0.7$ leads to an additional reduction of the storage, $G^{\prime}$, and loss, $G^{\prime\prime}$, modulus, yield strain $\gamma_{\mathrm{y}}$ and stress $\sigma_{\mathrm{y}}$, which indicates that the glass still becomes mechanically weaker.
\begin{figure}[t]
\centering
  \includegraphics[scale=0.25]{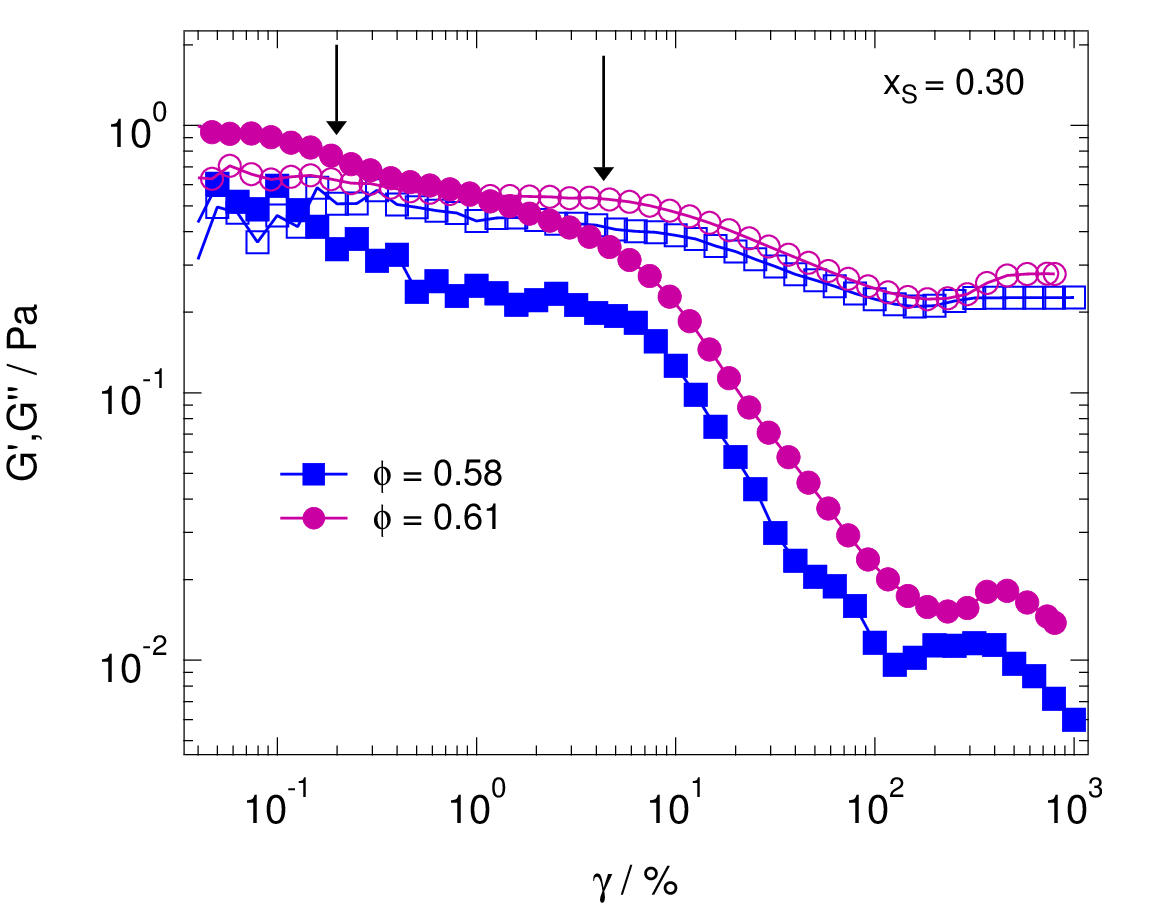}
  \caption{Storage, $G^{\prime}$, (full symbols) and loss, $G^{\prime\prime}$, (open symbols) moduli as a function of the strain amplitude $\gamma$ obtained in DSS measurements for $\delta = 0.20$, a relative volume fraction of small spheres $x_{\mathrm{S}} = 0.3$ and the total volume fraction $\phi=0.61$ (\magenta{{\Large{${\circ}$}}}) and $\phi=0.58$ (\blue{$\Box$)}. Arrows indicate the two yielding points observed for the sample with $\phi=0.61$.}
  \label{fig2:xs03}
\end{figure}
For $x_{\mathrm{S}} \ge 0.7$, comparable effects are found for $\phi=0.61$ and $0.58$.


This is different for  $x_{\mathrm{S}} < 0.7$.
For the higher total volume fraction $\phi=0.61$, the samples with $x_{\mathrm{S}}=0.5$ and 0.3 have a much smaller $G^{\prime}$ which, however, is still slightly larger than $G^{\prime\prime}$ and the samples hence show a weak solid-like response in the linear viscoelastic regime (Fig.~\ref{fig2:dss_diff_xs}a). 
This is consistent with $\gamma_{\mathrm{y}}$ and $\sigma_{\mathrm{y}}$ values which are more than one and almost three orders of magnitude smaller, respectively, than typical values of one-component hard-sphere glasses at the same total volume fraction. 
Hence the samples become very brittle and may flow plastically at smaller strain amplitudes or stresses.
A closer inspection of the response of the sample with $\phi=0.61$ and $x_{\mathrm{S}} = 0.3$ reveals a particularly interesting strain amplitude ($\gamma$) dependence of the moduli (Fig.~\ref{fig2:xs03}).
The linear response ends already at $\gamma \approx 0.2$\% (Fig.~\ref{fig2:xs03}, arrow on the left), beyond which $G^{\prime}$ decreases smoothly up to $\gamma \approx 4$\%, where it shows a kink and subsequently decreases with a power-law, while $G^{\prime\prime}$ shows a small maximum (Fig.~\ref{fig2:xs03}, arrow on the right). 
This response suggests the presence of two length scales, most likely associated with the small and large spheres, which both contribute to the yielding of the system at this $x_{\mathrm{S}}$. 
The first yielding at small strains $\gamma \approx 0.2$\% might correspond to plastic rearrangements of cages formed by small spheres. Cage distortion and yielding might be facilitated by the shear-induced interaction with the large spheres, i.e.~contact forces between large and small spheres.  
Once these cages are rearranged, the system is still prevented from flowing by the cages of large spheres which are only slightly deformed. 
At strains of about 4\% the cages of large spheres deform and the system starts to flow.
The ratio between the two yield strains, $0.2/4\simeq 0.04$, corresponds to $\delta/4$ which suggests a non trivial scaling of the yield strains with the cage size (which would give a factor $\delta$).
This finding could also result from the moderate polydispersity of the small spheres, which implies a distribution of the effective size ratio and accelerates the dynamics,\cite{schoepe:2007} and could contribute to smear out the double yielding phenomenon.
A two-steps yielding behavior has also been observed for attractive glasses and gels.\cite{pham06,pham08,laurati:jor:2011,koumakis:soft:2011}
Compared to $\phi=0.61$, for the lower total volume fraction $\phi=0.58$ decreasing the relative volume fraction of small spheres to $x_{\mathrm{S}}=0.5$ has an even stronger effect (Fig.~\ref{fig2:dss_diff_xs}b).
Within the whole examined range of strain amplitudes, $G^{\prime\prime}> G^{\prime}$ implying fluid-like behavior.
Thus, the glass is melted.
Fluid-like behavior in the whole range of measured $\gamma$ is also observed for $x_{\mathrm{S}} = 0.3$, with the response being similar to that obtained for $\phi=0.61$, except for the smallest $\gamma$ (Fig.~\ref{fig2:xs03}). Samples showing fluid-like behavior ($x_{\mathrm{S}} = 0.3$, 0.5) do not present a finite value of yield strain and stress, corresponding to the missing circles in Fig.~\ref{fig2:dss_diff_xs}b. 
The melting of the glass is caused by the larger free volume fraction created by the presence of small spheres, as will be discussed in more detail later.
This is similar to the behaviour of one-component systems when $\phi$ is decreased below the glass transition.

\begin{figure}[h]
\centering
\vspace{-0.5cm}
   \includegraphics[scale=0.4]{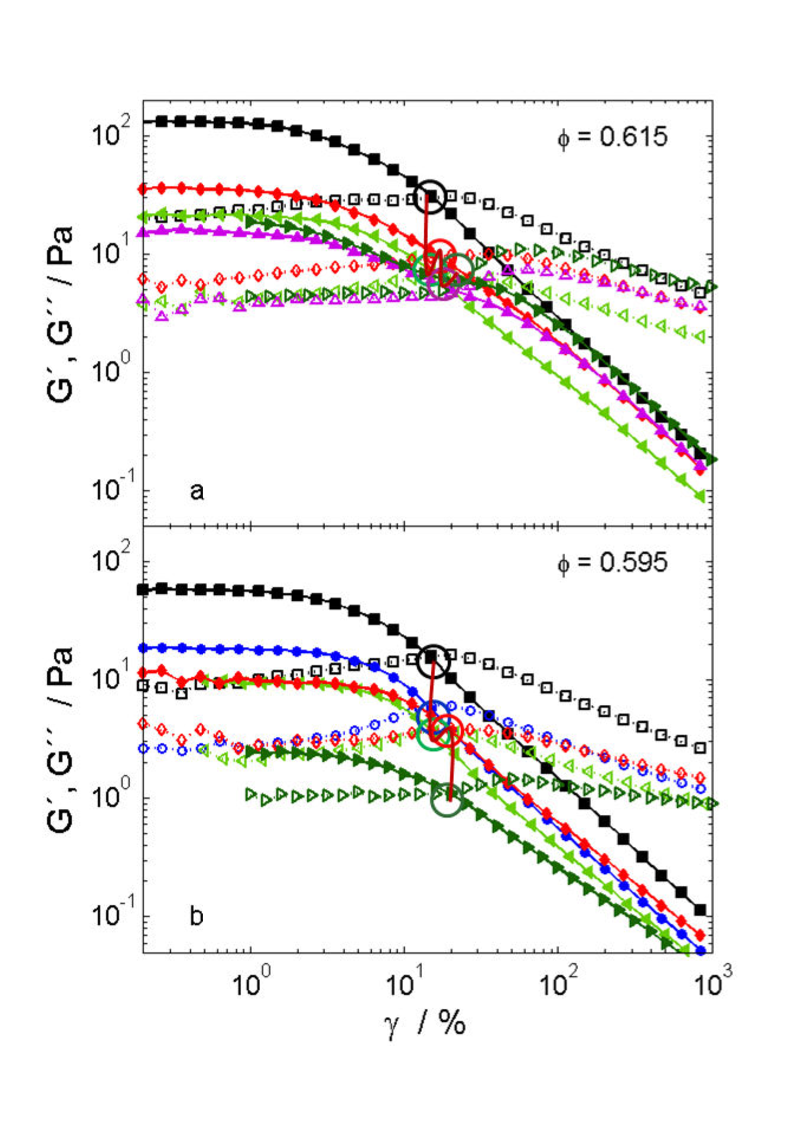}
	\vspace{-1cm}
  \caption{(a,b) Storage, $G^{\prime}$, (full symbols) and loss, $G^{\prime\prime}$, (open symbols) moduli as a function of strain amplitude $\gamma$ obtained in DSS measurements. The size ratio $\delta=0.38$, (a) the total volume fraction $\phi=0.615$ and the relative volume fraction of small particles $x_{\mathrm{S}}= 0.0$ (\olivegreen{$\triangleright$}), 0.08 (\purple{$\triangle$}), 0.25 (\red{$\Diamond$}), 0.5 (\green{$\triangleleft$}), 1.0 ($\Box$) and (b) $\phi=0.595$ and $x_{\mathrm{S}}=0.0$ (\olivegreen{$\triangleright$}), 0.25 (\red{$\Diamond$}), 0.5 (\green{$\triangleleft$}), 0.75 (\blue{{\Large ${\circ}$}}) and 1.0 ($\Box$). The frequency $\omega = 1$~rad/s. Circles indicate the yield strains $\gamma_{\mathrm{y}}$ and the red solid line their $x_{\mathrm{S}}$-dependence.
  }
  \label{fig2bis:dss_diff_xs}
\end{figure}   


Finally, the samples at both total volume fractions show the response of a weak solid for $x_{\mathrm{S}}=0.1$.
For $\phi = 0.61$ the storage modulus $G^{\prime}$ is further reduced and becomes similar to $G^{\prime\prime}$, indicating the proximity of a transition to the fluid state. On the other hand the yield strain $\gamma_{\mathrm{y}}$ and stress $\sigma_{\mathrm{y}}$ are slightly increased (Fig.~\ref{fig2:dss_diff_xs}).
In contrast, for $\phi = 0.58$ the response again changes qualitatively, which implies a reentrant behavior; the melting and re-formation of a solid glass state as the fraction of small spheres is reduced.


\begin{figure}[!h]
\centering
   \includegraphics[scale=0.38]{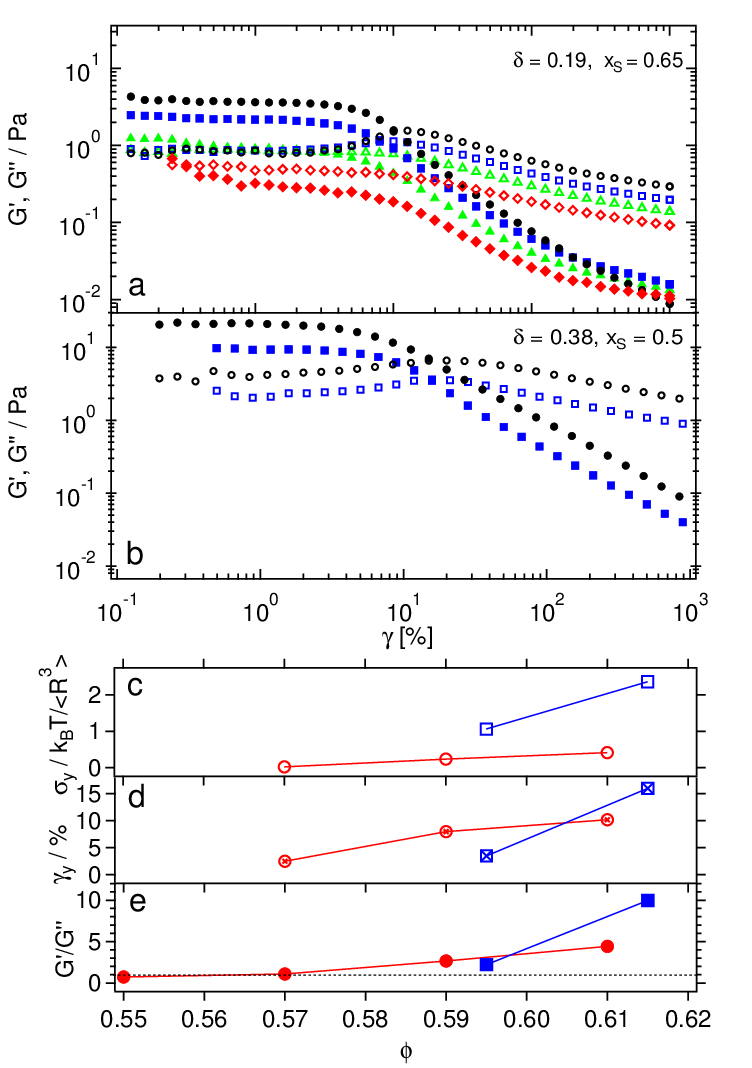}
  \caption{(a) Storage, $G^{\prime}$, (full symbols) and loss, $G^{\prime\prime}$, (open symbols) moduli as a function of strain amplitude $\gamma$ obtained in DSS measurements for (a) The size ratio $\delta = 0.19$, the relative volume fraction of small particles $x_{\mathrm{S}}=0.65$, total volume fractions $\phi=0.61$ ({\Large ${\circ}$}), $0.59$ (\blue{$\Box$}), $0.57$ (\green{$\triangle$}) and $0.55$ (\red{$\Diamond$}). (b) $\delta = 0.38$, $x_{\mathrm{S}} = 0.5$, $\phi= 0.615$ ({\Large ${\circ}$}) and $0.595$ (\blue{$\Box$}). Frequency  $\omega = 1$~rad/s, corresponding to $Pe_{\omega}=8.99\times10^{-2}$ for $\delta = 0.19$ and $Pe_{\omega}=8.85\times10^{-2}$ for $\delta = 0.38$. (c) Yield stress $\sigma_y$, (d) yield strain $\gamma_y$ and (e) ratio $G^{\prime}/G^{\prime\prime}$ in the linear viscoelastic regime ($\gamma = 0.5$\% and 1\% for $\delta = 0.2$ and 0.38, respectively), as a function of total volume fraction $\phi$ for samples of plot (a) ({\Large{\red{$\circ$}}}) and (b) (\blue{$\Box$}).}
  \label{fig4bis:dss_diff_phi}
\end{figure}

A second size ratio, $\delta = 0.38$, was investigated also at two total volume fractions $\phi = 0.595$ and 0.615 and different relative volume fractions of small particles $x_{\mathrm{S}}$ (Fig.~\ref{fig2bis:dss_diff_xs}). 
Starting from the one-component systems and increasing the amount of the second component, the storage modulus $G^{\prime}$ decreases in the linear viscoelastic regime indicating a softening of the glass, similar to the findings with $\delta = 0.20$ (Fig.~\ref{figure3:trends}a).
However, in the case of $\delta=0.38$, the minimum of $G^{\prime}$ is located at $x_{\mathrm{S}} \approx 0.5$ for both $\phi$. 
Note that in terms of the relative number of small spheres $\xi_S = n_L/(n_S+n_L)=x_S[\delta^3+x_S(1-\delta^3)]^{-1}$, where $n_{\mathrm{S}}$ and $n_{\mathrm{L}}$ are the number densities of small and large spheres, respectively, the minimum of $G^{\prime}$ is found for both size ratios at values of $\xi_S > 0.85$. 
Furthermore, the minimum in the $x_{\mathrm{S}}$-dependence is much weaker for the yield stress $\sigma_{\mathrm{y}}$ and absent for the yield strain $\gamma_{\mathrm{y}}$ (Fig.~\ref{figure3:trends}b,c).
Thus, no melting of the glass is observed for $\delta = 0.38$. 


Having studied the rheological response as a function of the relative volume fraction of small spheres $x_{\mathrm{S}}$, we now turn to the dependence on the total volume fraction $\phi$ for constant $x_{\mathrm{S}}=0.65$ ($\delta=0.19$) and 0.5 ($\delta=0.38$) (Fig.~\ref{fig4bis:dss_diff_phi}).
With decreasing $\phi$, the storage modulus $G^{\prime}$ decreases in the linear regime and approaches the loss modulus $G^{\prime\prime}$ (Fig.~\ref{fig4bis:dss_diff_phi}a,b, e, which shows the ratio $G^{\prime}/G^{\prime\prime}$).
Thus, with decreasing $\phi$, the solid-like response becomes weaker.
This is particularly pronounced for $\delta = 0.19$, which shows a fluid-like response for $\phi=0.55$, that is $G^{\prime\prime} > G^{\prime}$ in the linear viscoelastic regime (Fig.~\ref{fig4bis:dss_diff_phi} a,e).
The yield point, i.e.~the yield strain $\gamma_{\mathrm{y}}$ and stress $\sigma_{\mathrm{y}}$, decreases with decreasing $\phi$ for both values of $\delta$ and, for $\delta=0.19$ disappears at $\phi=0.55$, i.e.~the sample becomes a fluid (Fig.~\ref{fig4bis:dss_diff_phi}c,d).
This is consistent with the response of one-component systems, whose yield strain $\gamma_{\mathrm{y}}$ also decreases with decreasing $\phi$ until a transition to a fluid occurs.\cite{koumakis:2012,petekidis02,petekidis03}

\begin{figure}[t]
\centering
   \includegraphics[scale=0.45]{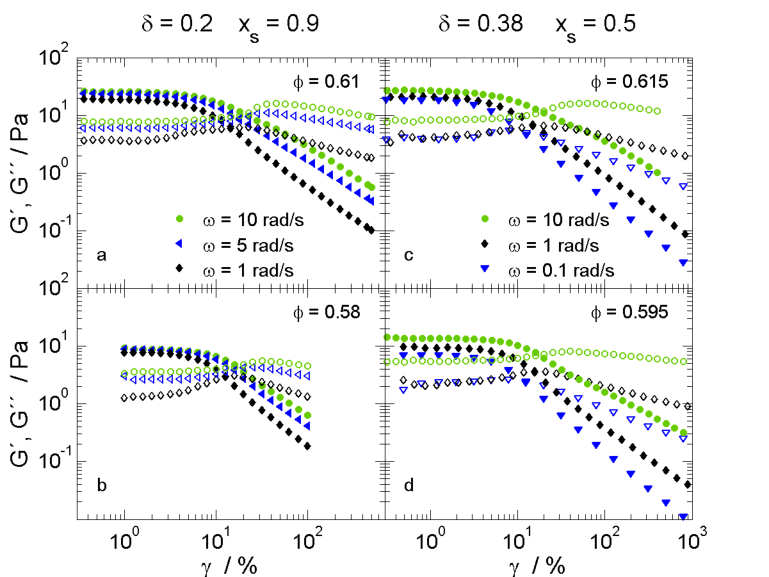}
  \caption{Storage, $G^{\prime}$, (full symbols) and loss, $G^{\prime\prime}$, (open symbols) moduli as a function of strain amplitude $\gamma$ obtained in DSS measurements. (left) The size ratio $\delta = 0.20$, total volume fraction (top) $\phi=0.61$ and (bottom) $0.58$, relative volume fraction of small spheres $x_{\mathrm{S}}=$ 0.9, and frequencies  $\omega = 1$~rad/s ($\Diamond$), 5~rad/s (\blue{$\triangleleft$}) and 10~rad/s (\green{{\Large ${\circ}$}}). (right) $\delta = 0.38$, (top) $\phi = 0.615$ and (bottom) $0.595$, $x_{\mathrm{S}}=0.5$, $\omega = 0.1$~rad/s (\blue{$\triangledown$}), 1~rad/s ($\Diamond$) and 10~rad/s (\green{{\Large ${\circ}$}}).}
  \label{fig4:dss_diff_omega}
\end{figure}

The decrease in $\gamma_{\mathrm{y}}$ is attributed to the fact that, upon decreasing $\phi$, the cages become larger and looser and thus increasingly smaller distortions of the cages are sufficient to allow the particles to escape through Brownian motion.
Finally, in the fluid phase ($\phi=0.55$), particles can leave the cage even in the absence of shear.
The sample with $\delta = 0.19$, $x_{\mathrm{S}}=0.65$ and $\phi = 0.57$ shows a dependence of $G^{\prime}$ and $G^{\prime\prime}$ on the strain amplitude $\gamma$ similar to that of the sample with $\delta=0.20$, $x_{\mathrm{S}} = 0.3$ and $\phi = 0.61$ (Fig.~\ref{fig2:xs03}), which again suggests the presence of two yielding points. Note  that this sample is a dense, slowly relaxing fluid and not a glass, according to the frequency dependence of the linear viscoelastic moduli (data not shown).  Nevertheless, the similarity of the response of the two samples suggests that a glass state similar to that of $\delta=0.20$, $x_{\mathrm{S}} = 0.3$ and $\phi = 0.61$, i.e.~characterized by a double yielding process and caging on two lenght scales, might be obtained at slightly larger $\phi$ for $x_{\mathrm{S}}=0.65$. This is in agreement with MCT predictions, where a transition from a glass characterized by caging on one length scale (that of the small spheres) at high $\phi$, to a glass characterized by caging on two length scales at lower $\phi$, and successive melting of this glass with further decreasing $\phi$, is expected along a line of constant $x_{\mathrm{S}}$.\cite{voigtmann:epl:2011,amokrane:2012} 

The results of the DSS measurements show a slight dependence on frequency (Fig.~\ref{fig4:dss_diff_omega}).
In the linear viscoelastic regime, the storage modulus $G^{\prime}$ increases with increasing frequency $\omega$, in agreement with results of our Dynamic Frequency Sweep (DFS) measurements (Fig.~\ref{fig1:phi_adjust_rheo}) and as discussed in more detail elsewhere.\cite{sentjabrskaya_linear_2012}
With increasing frequency $\omega$, the probed times decrease and are progressively shorter than the structural relaxation time.
This leads to an increasingly more elastic response.
Also the yield strain $\gamma_{\mathrm{y}}$ and stress $\sigma_{\mathrm{y}}$ increase with increasing $Pe_{\omega}$ (Fig.~\ref{figure3:trends}b,c).
This is similar to the behavior of one-component colloidal glasses \cite{petekidis02,petekidis03,pham06,pham08} and can be understood as follows.
Shear-induced cage deformation facilitates the escape of particles from their cage through Brownian motion, which results in yielding.\cite{nick:prl:12}
In oscillatory shear, the maximum cage deformation is achieved at the largest excursion.
In the vicinity of this point a particle is most likely to escape from the cage by Brownian motion. 
With increasing frequency, the particles spend less time at the maximum (but more frequently) and are therefore less likely to escape because the escape probability depends rather on the balance between the residence time at the maximum and the Brownian time than on the attempt rate.\cite{smith:pre:2007,laurati:jor:2011,conrad:421}. The reduced escape probability must be compensated by a larger cage deformation.
Thus, with increasing frequency $\omega$, a larger strain and stress will be required, and hence stored, before the cage breaks.\\


\begin{figure}[h]
\centering
  \includegraphics[scale=0.62]{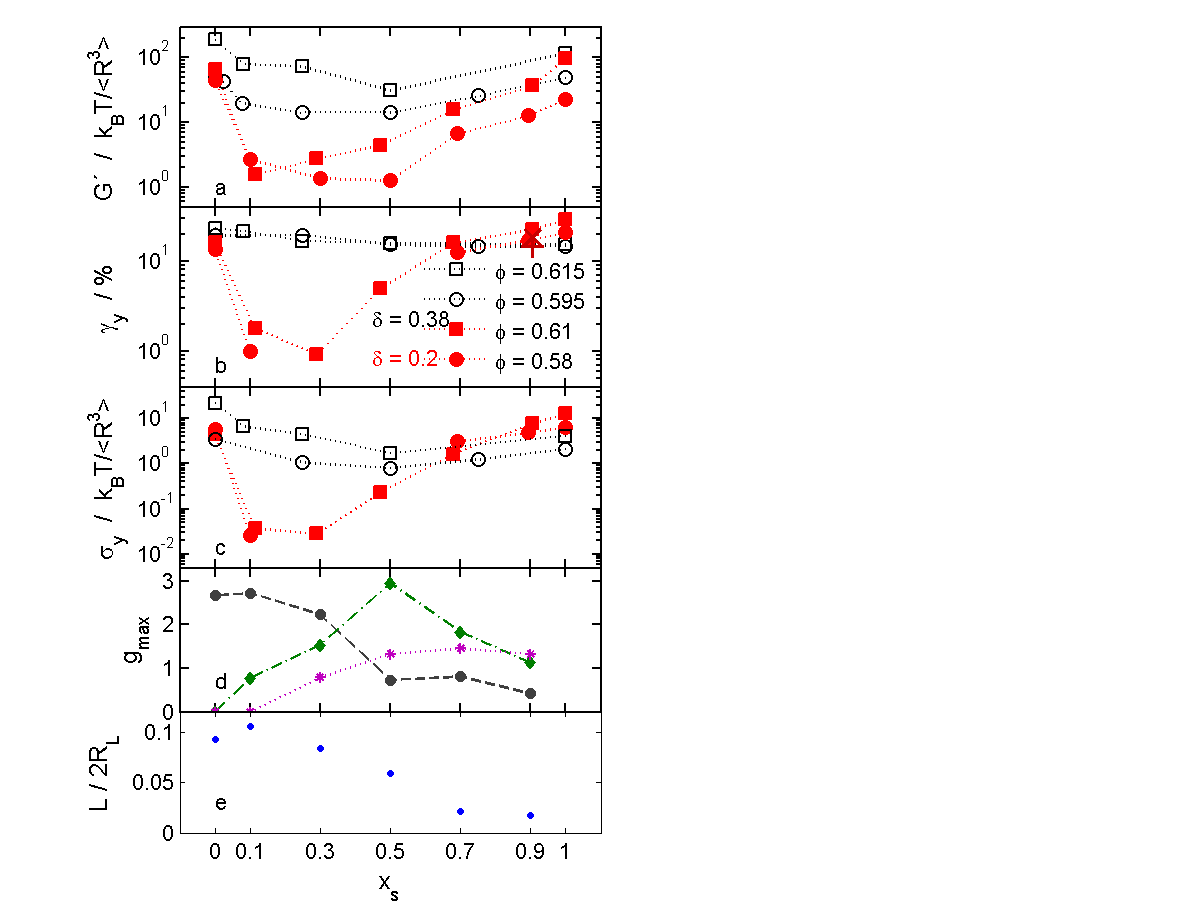}
  \caption{(a) Storage modulus $G^{\prime}$ in the linear viscoelastic regime ($\gamma = 0.2$\%), (b) yield strain $\gamma_{\mathrm{y}}$ estimated from the crossing point of $G^{\prime}$ and $G^{\prime\prime}$ and (c) corresponding yield stress $\sigma_{\mathrm{y}}$ as a function of the relative volume fraction of small particles $x_{\mathrm{S}}$ for samples with size ratio $\delta=0.20$ and total volume fraction $\phi=0.61$ (\red{$\blacksquare$}) and $\phi=0.58$ (\red{{\Large ${\bullet}$}}) and $Pe_{\omega}=5.55~10^{-1}$, and $\delta=0.38$, $\phi=0.615$ ($\Box$) and $\phi=0.595$ ({\Large ${\circ}$}) and $\omega = 1$ rad/s. (b) also contains results for $Pe_{\omega}=2.75~10^{-1}$ (\red{$\times$}) and $Pe_{\omega}=5.55~10^{-2}$ (\red{+}) for the sample with $x_{\mathrm{S}}=0.9$ and $\phi=0.61$. (d) Height of the maxima of the pair distribution function $g(r)$, $g_\mathrm{max}$, corresponding to $r=2R_L$ (${\Large{\circ}}$),  $r=2(R_L+R_S)$ (\green{$\blacklozenge$}) and $r=2(R_L+2R_S)$ (\magenta{*}) and (e) localisation length $L$ extracted from plateaus of mean squared displacements as a function of $x_{\mathrm{S}}$, for samples with $\delta = 0.2$ and $\phi = 0.61$.\cite{sentjabrskaya_linear_2012_2,sentjabrskaya_linear_2012} Error bars are smaller than the symbols in all plots.}
  \label{figure3:trends}
\end{figure} 
 
Our findings are summarized in Fig.~\ref{figure3:trends}.
For a given total volume fraction $\phi$, adding a second component to the one-component systems results in a weaker elastic response. 
For $\delta = 0.20$, the glass softens particularly strongly and, if the sample is sufficiently close to the glass transition (here $\phi = 0.58$), even melts, that is shows a fluid-like response. 
This reduction in $G^{\prime}$ is not symmetric with respect to the one-component systems, but is more pronounced for glasses mainly consisting of large spheres to which a small amount of small spheres has been added.
This is evident when comparing, for example, $G^{\prime}$ for samples with $x_{\mathrm{S}} = 0.1$ and 0.9.
This asymmetry might, however, be due to the choice of the control parameter, here the relative volume fraction of small particles $x_{\mathrm{S}}$.
Instead, one could use the relative number of small spheres, $\xi_{\mathrm{S}}$.
Hence $x_{\mathrm{S}} = 0.1$ corresponds to $\xi_{\mathrm{S}} = 0.93$ while $x_{\mathrm{S}} = 0.9$ implies a relative number of large spheres of only $\xi_{\mathrm{L}} =  8.9\times10^{-4}$.
This might explain why for $x_{\mathrm{S}} = 0.9$ the cage of small spheres is not significantly affected by the small number of large spheres.
In contrast, for $x_{\mathrm{S}} = 0.1$ a large number of small spheres has to be accommodated by the large spheres, which is likely to induce a significant cage deformation and to result in a significant softening.
This is supported by confocal microscopy measurements of the structure and dynamics of the large spheres,\cite{sentjabrskaya_linear_2012_2,sentjabrskaya_linear_2012} which are summarized in Fig.\ref{figure3:trends}d,e. 
Already at $x_S = 0.1$, the pair distribution function $g(r)$ does not only show a peak at $r=2R_{\mathrm{L}}$, but also a shoulder at $r=2(R_{\mathrm{L}}+R_{\mathrm{S}})$ indicating that the cage of large spheres is deformed and that a significant fraction of large particles is separated by small particles. This cage deformation leads to a slight increase in the particle localisation length extracted from the plateau of mean-squared displacements, but the dynamics of the system is still arrested.\cite{sentjabrskaya_linear_2012_2,sentjabrskaya_linear_2012} The reduced localisation is thought to be responsible for the strong decrease in yield strain. Rearrangement of the cage of large spheres becomes even more pronounced as $x_{\mathrm{S}}$ is increased to $0.3$ and 0.5, as demonstrated by the increasingly larger reduction of the peak at $r=2R_{\mathrm{L}}$ and the corresponding increase at $r=2(R_{\mathrm{L}}+R_{\mathrm{S}})$ as well as the appearance of additional peaks at distances $r=2(R_{\mathrm{L}}+nR_{\mathrm{S}})$, with $n$ an integer number. For these $x_{\mathrm{S}}$, the dynamics show diffusion at $\phi=0.58$ and sub-diffusion at $\phi = 0.61$ with a decreasing localization length suggesting that large particles start to be localized more tightly by small spheres.\cite{sentjabrskaya_linear_2012_2,sentjabrskaya_linear_2012} 
During this process of cage rearrangement, for $\phi=0.58$ the elasticity decreases and the yield strain $\gamma_y$ disappears due to the melting of the glass, while at $\phi=0.61$ both $G^\prime$ and $\gamma_y$ start to increase again significantly above $x_S=0.3$ possibly due to the emergent caging and localisation of large spheres by small spheres. For $x_S > 0.5$ the localisation in cages of small spheres, i.e.~the transition to a different glass state, is accomplished: Large particles represent a dilute phase in a dense matrix of small spheres and are localised on distances which are about a factor $\delta=0.2$ smaller than at $x_S = 0$ and their dynamics are again arrested.\cite{sentjabrskaya_linear_2012_2,sentjabrskaya_linear_2012} 
 The tighter localisation and dynamical arrest induce an increased $G^\prime$ and  $\gamma_y$ towards the values of the one-component glass of small spheres. A pronounced effect of size and mixing ratios on the structure and dynamics of the glass was also reported for 2D colloidal glass formers.\cite{koenig_epje05,onuki_pre07,yunker_prl10,ebert:2008,bonales:2012} In particular, changes in the relative content of the small component and the size ratio have been reported to have pronounced effects on the dynamics.\cite{onuki_pre07,yunker_prl10}
 
 \begin{figure}[h]
\centering
   \includegraphics[scale=0.38]{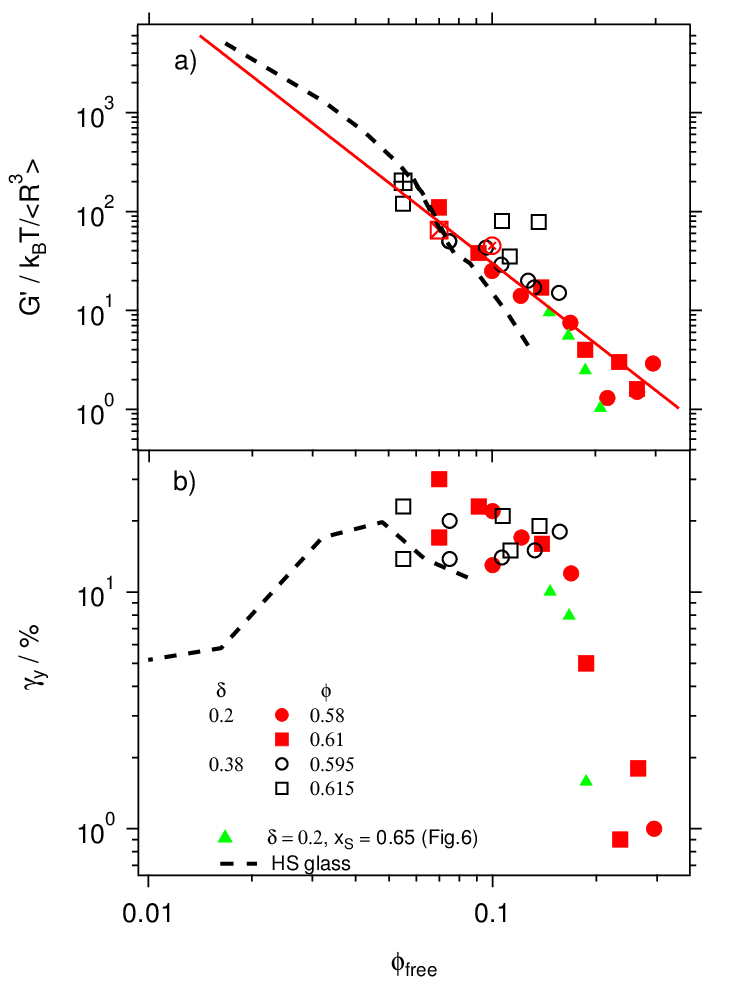}
  \caption{(a) Storage modulus $G^{\prime}$ in the linear viscoelastic regime and (b) yield strain $\gamma_{\mathrm{y}}$ as a function of the free volume $\phi_{\mathrm{free}}$ for the same samples as presented in Fig.~\ref{figure3:trends}, and for samples of Fig.~\ref{fig4bis:dss_diff_phi} (\green{$\blacktriangle$}). The red line in (a) shows a power-law fit $G^{\prime}\langle R^3\rangle/k_BT\sim\phi_{\mathrm{free}}^{-p}$, with p$\approx 3$. The dashed lines represent data of one-component hard-sphere glasses.\cite{koumakis:2012}}
  \label{fig_rcp}
\end{figure} 
 
For $\delta = 0.38$, the softening is less pronounced and no melting is observed.
Moreover, the dependence of $G^{\prime}$ on $x_{\mathrm{S}}$ is more symmetrical with respect to the one-component systems.
The smaller effect is attributed to the fact that the small particles have a reduced ability to occupy the interstitial space between the large particles at this size ratio. The critical value $\delta_c$ at which the small spheres cannot fill the space in between two large ones in a dense packing of large spheres can be estimated: In a group of 9 spheres arranged as in a body-centred cubic lattice and in contact with each other, the centers of two spheres along a face diagonal are separated by $2\sqrt{2}R_{\mathrm{L}}$ and a small sphere can fill the space left in between the large spheres if $R_{\mathrm{S}} \leq (\sqrt{2}-1)R_{\mathrm{L}} \approx 0.41 R_{\mathrm{L}}$, i.e. $\delta \leq \delta_{\mathrm{c}} \approx 0.41$, which is comparable to $\delta = 0.38$. Although in the glass states considered here, ordered configurations are not expected, the size of the void space might be similar.
Thus the cage itself, i.e.~the first neighbo\s{u}r shell, is not expected to be rearranged significantly and the softening hence appears to be caused by the heterogeneity of the cage on an intermediate length scale rather than a more efficient packing.
The weaker cage deformation induced by the smaller packing ability at $\delta = 0.38$ can also explain the weaker reduction of the yield strain and stress observed at intermediate mixing ratios for this $\delta$.

Instead of the relative volume, $x_{\mathrm{S}}$, or number, $\xi_{\mathrm{S}}$, fraction of small particles, we now consider the distance to the two limiting volume fractions of the glass state, corresponding to the glass transition and random close packing.
Mode Coupling Theory (MCT) predicts \cite{voigtmann:epl:2011} that in mixtures the glass transition is shifted to higher total volume fractions.
For example, for the size ratio $\delta=0.38$ the maximum volume fraction for the glass transition is expected at $x_{\mathrm{S}} \approx 0.4$, which is consistent with the occurrence of maximum softening in our experiments.
The shift of the glass transition could be related to the addition of small particles with their larger mobility.
This might favour structural rearrangements of the large spheres through collective motions and lead to a glass with a reduced elasticity, i.e. $G^\prime$.
In mixtures, MCT predicts qualitative changes of the relative particle mobilities, associated with different glass states.
In addition, the more efficient packing in mixtures results in an increased total volume fraction at random close packing, $\phi_{\mathrm{RCP}}$. 
Theoretical predictions for $\phi_{\mathrm{RCP}}$ are available for binary mixtures of monodisperse hard spheres, with different size ratios $\delta$ and mixing ratios, i.e.~$x_{\mathrm{S}}$.\cite{biazzo:2009,farr:2009}
Based on the predictions for $\phi_{\mathrm{RCP}}$, we calculate the available free volume $\phi_{\mathrm{free}} = \phi_{\mathrm{RCP}}-\phi$ as a function of $x_{\mathrm{S}}$ and $\delta$.
(Predictions for $\delta = 0.17$ and 0.39 are used for the experimental $\delta = 0.20$ and 0.38, respectively.) Note that predicted values of $\phi_{\mathrm{RCP}}$ where shifted by the difference between the value of $\phi_\mathrm{RCP}$ in the monodisperse case ($\phi_\mathrm{RCP} = 0.4$) and the experimental values of $\phi_\mathrm{RCP}$ ($\phi_\mathrm{RCP} = 0.68$ for $\delta = 0.2$ and $\phi_\mathrm{RCP} = 0.67$ for $\delta = 0.38$, section \ref{samples}).
With decreasing free volume $\phi_{\mathrm{free}}$, that is toward random close packing, the storage modulus $G^{\prime}$ is found to increase  (Fig.~\ref{fig_rcp}).
The dependence of $G^{\prime}$ on $\phi_{\mathrm{free}}$ indicates a common behavior for all $\delta$ and $\phi$ investigated and can be approximately described by a power-law dependence $G^{\prime}\langle R^3\rangle/k_BT\sim(\phi_{\mathrm{free}})^{-p}$, with $p \approx 3$. A similar power-law dependence is observed for one-component hard-sphere systems up to $\phi_\mathrm{free}\leq 0.1$(Fig.~~\ref{fig_rcp}, dashed line).\cite{petekidis02,petekidis03,koumakis:2008,koumakis:2012} At larger values of $\phi_{\mathrm{free}}$ the one-component system shows a sharper decay.\\

We now consider the dependence of the yield point on the free volume $\phi_{\mathrm{free}}$.
The $x_{\mathrm{S}}$-dependence of the yield strain $\gamma_{\mathrm{y}}$ and stress $\sigma_{\mathrm{y}}$ is quite different for the two size ratios (Fig.~\ref{figure3:trends}b,c). 
In particular, both, $\gamma_{\mathrm{y}}$ and $\sigma_{\mathrm{y}}$, show a much weaker dependence on $x_{\mathrm{S}}$ for $\delta = 0.38$ than $\delta = 0.2$.
This can also be linked to the free volume available for structural rearrangements.
The dependence of  $\gamma_{\mathrm{y}}$ on $\phi_{\mathrm{free}}$ (Fig.~\ref{fig_rcp}b) indicates that toward small free volumes, the yield strain saturates at an approximately constant value $\gamma_{\mathrm{y}} \approx 20$\%, which agrees with the yield strain observed in one-component glasses.\cite{pham06,pham08,koumakis:2008,koumakis:2012} At smaller values of $\phi_{\mathrm{free}}$, i.e.~very close to RCP, which are not reached here, in the one-component systems the yield strain decreases (Fig.\ref{fig_rcp}).   
In contrast, toward large $\phi_{\mathrm{free}} > 0.1$ a strong decrease of $\gamma_{\mathrm{y}}$ is observed (for samples with $\delta = 0.20$ since only they reach large enough $\phi_{\mathrm{free}}$ due to their large $\phi_{\mathrm{RCP}}$). 
This decrease indicates that if a sufficiently large free volume, i.e.~a sufficiently loose packing, is present, significant structural rearrangements can be induced by small strains. 
Their strong decrease of the yield strain is observed for samples in which the small spheres occupy the free space between the large spheres.
The intercalation of small spheres in between large spheres possibly induces a strong deformation of the cage. Similar effects have been observed in mixtures of star polymers with significant size disparity.\cite{mayer:2008} 
This supports our previous finding that yielding is not only facilitated by the increase of free volume but also by structural heterogeneities leading to cage deformation.
Interestingly, the strong decrease in the yield strain $\gamma_{\mathrm{y}}$ for $\phi_{\mathrm{free}} > 0.1$ is not observed in one-component systems,\cite{petekidis02,petekidis03,koumakis:2012} since in this regime the glass is melted. 
This is also consistent with $G^{\prime}$ sharply decreasing for $\phi_\mathrm{free}\geq 0.1$ for the one-component system (Fig.~\ref{fig_rcp}a, dashed line). We speculate that in the glass state $G^{\prime}\langle R^3\rangle/k_BT\sim(\phi_{\mathrm{free}})^{-p}$ with $p\approx 3$ for one and two-component systems. The slight shift between our system (red line) and the previous one-component data (dashed line) is due to different interactions mediated by different solvents.\cite{Poon/Weeks/Royall}
These findings show that at large values of $\phi_{\mathrm{free}}$, a glass can still be formed in the mixture (possibly due to attractions) while a dense fluid is observed in the one-component system.  
\section{Conclusions}

The linear and non-linear response to oscillatory shear has been studied in concentrated binary hard-sphere mixtures with large size disparities, $\delta \approx 0.20$ and 0.38. 
In the linear regime, the response of mixtures is softer than that of the corresponding one-component systems at the same total volume fraction $\phi$, as demonstrated by the smaller normalised storage modulus $G^{\prime}$.
The softening is associated with a shift of random close packing to larger total volume fractions, and thus a larger free volume fraction $\phi_{\mathrm{free}}$, which results from the more efficient packing in two-component systems.\cite{biazzo:2009}
Pronounced softening occurs for the size ratio $\delta = 0.20$ and for samples containing a majority of large spheres ($x_{\mathrm{S}} \lesssim 0.5$).
This indicates that softening is not only a result of an increased free volume $\phi_{\mathrm{free}}$ but also of cage distortions  due to small particles filling the space between the large spheres.
In contrast, in the samples with a smaller size disparity ($\delta = 0.38$) and a majority of small spheres ($x_{\mathrm{S}} \gtrsim 0.5$), we can speculate that on average the cage structure should be poorly affected due to the reduced ability of the small component to fill space in between the large spheres, and heterogeneities are thus only introduced beyond the first neighbor shell, which results in a weaker softening of the glass.

In the non-linear regime, the more efficient packing in the mixtures affects the yielding behaviour. 
If the free volume $\phi_{\mathrm{free}}$ is only slightly increased, yielding is characterised by a one-step cage break-up, as in one-component systems.
With increasing free volume, yielding occurs at smaller deformations. 
Interestingly, at large values of the free volume, the presence of a small but finite yield strain indicates the persistence of a weak solid-like state in the mixtures, while at comparable free volume a one-component system melts. 
This occurs in systems where the small spheres can occupy the space in between the large spheres, which suggests that the intercalation of small spheres induces a strong deformation and loosening of the cage structure and thus contributes to the reduction of the yield strain.
Moreover, the yielding behaviour could be affected by a possible transition between different glass states, in particular if it is associated with the mobility of the small spheres, which could facilitate yielding through collective motions.
In addition to the one-step yielding behaviour, we also found indications of a more complex two-step yielding behavior for a sample\sw{s} with $x_{\mathrm{S}} = 0.3$, $\phi = 0.61$. 
The two steps could be linked to the two different length scales present in these samples, representing caging of small and large spheres, respectively.
While two length scales are present in all mixtures, in most samples one of the two dominates, rendering the second yielding insignificant.

\section{Acknowledgments}
We acknowledge support from the Deutsche Forschungsgemeinschaft through the FOR1394 Research unit and EU funding through the FP7-Infrastructures 'ESMI' (CP\&CSA-2010-262348). We also thank A. B. Schofield for the synthesis of the PMMA particles and N. Koumakis for his help at the initial stages of experiments at FORTH. We thank in addition Th. Voigtmann, K. J. Mutch, P. Chauduri, J. Horbach, M. Fuchs, R. Casta\~{n}eda Priego  and W. C. K. Poon for stimulating discussions.

\footnotesize{
\bibliography{paper_sentjabrskaya_nonlinear_resub} 

\providecommand*{\mcitethebibliography}{\thebibliography}
\csname @ifundefined\endcsname{endmcitethebibliography}
{\let\endmcitethebibliography\endthebibliography}{}
\begin{mcitethebibliography}{65}
\providecommand*{\natexlab}[1]{#1}
\providecommand*{\mciteSetBstSublistMode}[1]{}
\providecommand*{\mciteSetBstMaxWidthForm}[2]{}
\providecommand*{\mciteBstWouldAddEndPuncttrue}
  {\def\EndOfBibitem{\unskip.}}
\providecommand*{\mciteBstWouldAddEndPunctfalse}
  {\let\EndOfBibitem\relax}
\providecommand*{\mciteSetBstMidEndSepPunct}[3]{}
\providecommand*{\mciteSetBstSublistLabelBeginEnd}[3]{}
\providecommand*{\EndOfBibitem}{}
\mciteSetBstSublistMode{f}
\mciteSetBstMaxWidthForm{subitem}
{(\emph{\alph{mcitesubitemcount}})}
\mciteSetBstSublistLabelBeginEnd{\mcitemaxwidthsubitemform\space}
{\relax}{\relax}

\bibitem[Amokrane \emph{et~al.}(2005)Amokrane, Ayadim, and
  Malherbe]{amokrane:2005}
S.~Amokrane, A.~Ayadim and J.~Malherbe, \emph{J. Chem. Phys.}, 2005,
  \textbf{123}, 174508\relax
\mciteBstWouldAddEndPuncttrue
\mciteSetBstMidEndSepPunct{\mcitedefaultmidpunct}
{\mcitedefaultendpunct}{\mcitedefaultseppunct}\relax
\EndOfBibitem
\bibitem[Roth \emph{et~al.}(2000)Roth, Evans, and Dietrich]{roth:2000}
R.~Roth, R.~Evans and S.~Dietrich, \emph{Phys. Rev. E}, 2000, \textbf{62},
  5360\relax
\mciteBstWouldAddEndPuncttrue
\mciteSetBstMidEndSepPunct{\mcitedefaultmidpunct}
{\mcitedefaultendpunct}{\mcitedefaultseppunct}\relax
\EndOfBibitem
\bibitem[Ashton \emph{et~al.}(2011)Ashton, Wilding, Roth, and
  Evans]{ashton:2011}
D.~J. Ashton, N.~B. Wilding, R.~Roth and R.~Evans, \emph{Phys. Rev. E}, 2011,
  \textbf{84}, 061136\relax
\mciteBstWouldAddEndPuncttrue
\mciteSetBstMidEndSepPunct{\mcitedefaultmidpunct}
{\mcitedefaultendpunct}{\mcitedefaultseppunct}\relax
\EndOfBibitem
\bibitem[Dijkstra \emph{et~al.}(1999)Dijkstra, van Roij, and
  Evans]{dijkstra:99}
M.~Dijkstra, R.~van Roij and R.~Evans, \emph{Phys. Rev. E}, 1999, \textbf{59},
  5744\relax
\mciteBstWouldAddEndPuncttrue
\mciteSetBstMidEndSepPunct{\mcitedefaultmidpunct}
{\mcitedefaultendpunct}{\mcitedefaultseppunct}\relax
\EndOfBibitem
\bibitem[G{\"o}tzelmann \emph{et~al.}(1999)G{\"o}tzelmann, Roth, Dietrich,
  Dijkstra, and Evans]{goetzelmann:99}
B.~G{\"o}tzelmann, R.~Roth, S.~Dietrich, M.~Dijkstra and R.~Evans,
  \emph{Europhys. Lett.}, 1999, \textbf{47}, 398\relax
\mciteBstWouldAddEndPuncttrue
\mciteSetBstMidEndSepPunct{\mcitedefaultmidpunct}
{\mcitedefaultendpunct}{\mcitedefaultseppunct}\relax
\EndOfBibitem
\bibitem[Malherbe and Krauth(2007)]{malherbe:2007}
J.~G. Malherbe and W.~Krauth, \emph{Mol. Phys.}, 2007, \textbf{105}, 2393\relax
\mciteBstWouldAddEndPuncttrue
\mciteSetBstMidEndSepPunct{\mcitedefaultmidpunct}
{\mcitedefaultendpunct}{\mcitedefaultseppunct}\relax
\EndOfBibitem
\bibitem[{van Duijneveldt} \emph{et~al.}(1993){van Duijneveldt}, Heinen, and
  Lekkerkerker]{van_duijnevelt:93}
J.~S. {van Duijneveldt}, A.~W. Heinen and H.~N.~W. Lekkerkerker,
  \emph{Europhys. Lett.}, 1993, \textbf{21}, 369\relax
\mciteBstWouldAddEndPuncttrue
\mciteSetBstMidEndSepPunct{\mcitedefaultmidpunct}
{\mcitedefaultendpunct}{\mcitedefaultseppunct}\relax
\EndOfBibitem
\bibitem[Dinsmore \emph{et~al.}(1995)Dinsmore, Yodh, and Pine]{dinsmore:95}
A.~D. Dinsmore, A.~G. Yodh and D.~J. Pine, \emph{Phys. Rev. E}, 1995,
  \textbf{52}, 4045\relax
\mciteBstWouldAddEndPuncttrue
\mciteSetBstMidEndSepPunct{\mcitedefaultmidpunct}
{\mcitedefaultendpunct}{\mcitedefaultseppunct}\relax
\EndOfBibitem
\bibitem[Imhof and Dhont(1995)]{imhof:prl:95}
A.~Imhof and J.~K.~G. Dhont, \emph{Phys. Rev. Lett.}, 1995, \textbf{75},
  1662\relax
\mciteBstWouldAddEndPuncttrue
\mciteSetBstMidEndSepPunct{\mcitedefaultmidpunct}
{\mcitedefaultendpunct}{\mcitedefaultseppunct}\relax
\EndOfBibitem
\bibitem[Hunt \emph{et~al.}(2000)Hunt, Jardine, and Bartlett]{hunt:2000}
N.~Hunt, R.~Jardine and P.~Bartlett, \emph{Phys. Rev. E}, 2000, \textbf{62},
  900\relax
\mciteBstWouldAddEndPuncttrue
\mciteSetBstMidEndSepPunct{\mcitedefaultmidpunct}
{\mcitedefaultendpunct}{\mcitedefaultseppunct}\relax
\EndOfBibitem
\bibitem[Bartlett \emph{et~al.}(1992)Bartlett, Ottewill, and
  Pusey]{bartlett:92}
P.~Bartlett, R.~H. Ottewill and P.~N. Pusey, \emph{Phys. Rev. Lett.}, 1992,
  \textbf{68}, 3801\relax
\mciteBstWouldAddEndPuncttrue
\mciteSetBstMidEndSepPunct{\mcitedefaultmidpunct}
{\mcitedefaultendpunct}{\mcitedefaultseppunct}\relax
\EndOfBibitem
\bibitem[Cottin and Monson(1995)]{cottin:95}
X.~Cottin and P.~A. Monson, \emph{J. Chem. Phys.}, 1995, \textbf{102},
  3354\relax
\mciteBstWouldAddEndPuncttrue
\mciteSetBstMidEndSepPunct{\mcitedefaultmidpunct}
{\mcitedefaultendpunct}{\mcitedefaultseppunct}\relax
\EndOfBibitem
\bibitem[Schofield(2001)]{schofield:2001}
A.~B. Schofield, \emph{Phys. Rev. E}, 2001, \textbf{64}, 051403\relax
\mciteBstWouldAddEndPuncttrue
\mciteSetBstMidEndSepPunct{\mcitedefaultmidpunct}
{\mcitedefaultendpunct}{\mcitedefaultseppunct}\relax
\EndOfBibitem
\bibitem[Hynninen \emph{et~al.}(2009)Hynninen, Filion, and
  Dijkstra]{hynninen:09}
A.-P. Hynninen, L.~Filion and M.~Dijkstra, \emph{J. Chem. Phys.}, 2009,
  \textbf{131}, 064902\relax
\mciteBstWouldAddEndPuncttrue
\mciteSetBstMidEndSepPunct{\mcitedefaultmidpunct}
{\mcitedefaultendpunct}{\mcitedefaultseppunct}\relax
\EndOfBibitem
\bibitem[Dijkstra \emph{et~al.}(1998)Dijkstra, van Roij, and
  Evans]{dijkstra:98}
M.~Dijkstra, R.~van Roij and R.~Evans, \emph{Phys. Rev. Lett.}, 1998,
  \textbf{81}, 2268\relax
\mciteBstWouldAddEndPuncttrue
\mciteSetBstMidEndSepPunct{\mcitedefaultmidpunct}
{\mcitedefaultendpunct}{\mcitedefaultseppunct}\relax
\EndOfBibitem
\bibitem[G\"otze and {Th. Voigtmann}(2003)]{voigtmann:2003}
W.~G\"otze and {Th. Voigtmann}, \emph{Phys. Rev. E}, 2003, \textbf{67},
  021502\relax
\mciteBstWouldAddEndPuncttrue
\mciteSetBstMidEndSepPunct{\mcitedefaultmidpunct}
{\mcitedefaultendpunct}{\mcitedefaultseppunct}\relax
\EndOfBibitem
\bibitem[{Th. Voigtmann}(2011)]{voigtmann:epl:2011}
{Th. Voigtmann}, \emph{Eur. Phys. Lett.}, 2011, \textbf{96}, 36006\relax
\mciteBstWouldAddEndPuncttrue
\mciteSetBstMidEndSepPunct{\mcitedefaultmidpunct}
{\mcitedefaultendpunct}{\mcitedefaultseppunct}\relax
\EndOfBibitem
\bibitem[{Ph. Germain} and Amokrane(2009)]{germain:2009}
{Ph. Germain} and S.~Amokrane, \emph{Phys. Rev. Lett.}, 2009, \textbf{102},
  058301\relax
\mciteBstWouldAddEndPuncttrue
\mciteSetBstMidEndSepPunct{\mcitedefaultmidpunct}
{\mcitedefaultendpunct}{\mcitedefaultseppunct}\relax
\EndOfBibitem
\bibitem[{Tchangnwa Nya} \emph{et~al.}(2012){Tchangnwa Nya}, Ayadim, {Ph.
  Germain}, and Amokrane]{amokrane:2012}
F.~{Tchangnwa Nya}, A.~Ayadim, {Ph. Germain} and S.~Amokrane, \emph{J. Phys.:
  Condens. Matter}, 2012, \textbf{24}, 325106\relax
\mciteBstWouldAddEndPuncttrue
\mciteSetBstMidEndSepPunct{\mcitedefaultmidpunct}
{\mcitedefaultendpunct}{\mcitedefaultseppunct}\relax
\EndOfBibitem
\bibitem[Williams and van Megen(2001)]{williams:2001}
S.~R. Williams and W.~van Megen, \emph{Phys. Rev. E}, 2001, \textbf{64},
  041502\relax
\mciteBstWouldAddEndPuncttrue
\mciteSetBstMidEndSepPunct{\mcitedefaultmidpunct}
{\mcitedefaultendpunct}{\mcitedefaultseppunct}\relax
\EndOfBibitem
\bibitem[Rodriguez \emph{et~al.}(1992)Rodriguez, Kaler, and
  Wolfe]{rodriguez:92}
B.~E. Rodriguez, E.~W. Kaler and M.~S. Wolfe, \emph{Langmuir}, 1992,
  \textbf{8}, 2382\relax
\mciteBstWouldAddEndPuncttrue
\mciteSetBstMidEndSepPunct{\mcitedefaultmidpunct}
{\mcitedefaultendpunct}{\mcitedefaultseppunct}\relax
\EndOfBibitem
\bibitem[Foffi \emph{et~al.}(2003)Foffi, G\"otze, Sciortino, Tartaglia, and
  {Th. Voigtmann}]{foffi:2003}
G.~Foffi, W.~G\"otze, F.~Sciortino, P.~Tartaglia and {Th. Voigtmann},
  \emph{Phys. Rev. Lett.}, 2003, \textbf{91}, 085701\relax
\mciteBstWouldAddEndPuncttrue
\mciteSetBstMidEndSepPunct{\mcitedefaultmidpunct}
{\mcitedefaultendpunct}{\mcitedefaultseppunct}\relax
\EndOfBibitem
\bibitem[Pham \emph{et~al.}(2002)Pham, Puertas, Bergenholtz, Egelhaaf,
  Moussaid, Pusey, Schofield, Cates, Fuchs, and Poon]{pham02}
K.~N. Pham, A.~M. Puertas, J.~Bergenholtz, S.~U. Egelhaaf, A.~Moussaid, P.~N.
  Pusey, A.~B. Schofield, M.~Cates, M.~Fuchs and W.~C.~K. Poon, \emph{Science},
  2002, \textbf{296}, 104--106\relax
\mciteBstWouldAddEndPuncttrue
\mciteSetBstMidEndSepPunct{\mcitedefaultmidpunct}
{\mcitedefaultendpunct}{\mcitedefaultseppunct}\relax
\EndOfBibitem
\bibitem[Pham \emph{et~al.}(2004)Pham, Egelhaaf, Pusey, and Poon]{pham04}
K.~N. Pham, S.~U. Egelhaaf, P.~N. Pusey and W.~C.~K. Poon, \emph{Phys. Rev. E},
  2004, \textbf{69}, 011503\relax
\mciteBstWouldAddEndPuncttrue
\mciteSetBstMidEndSepPunct{\mcitedefaultmidpunct}
{\mcitedefaultendpunct}{\mcitedefaultseppunct}\relax
\EndOfBibitem
\bibitem[Woutersen and de~Kruif(1993)]{woutersen:93}
A.~T. J.~M. Woutersen and C.~G. de~Kruif, \emph{J. Rheol.}, 1993, \textbf{37},
  681\relax
\mciteBstWouldAddEndPuncttrue
\mciteSetBstMidEndSepPunct{\mcitedefaultmidpunct}
{\mcitedefaultendpunct}{\mcitedefaultseppunct}\relax
\EndOfBibitem
\bibitem[{D'Haene} and Mewis(1994)]{mewis:94}
P.~{D'Haene} and J.~Mewis, \emph{Rheol. Acta}, 1994, \textbf{33}, 165\relax
\mciteBstWouldAddEndPuncttrue
\mciteSetBstMidEndSepPunct{\mcitedefaultmidpunct}
{\mcitedefaultendpunct}{\mcitedefaultseppunct}\relax
\EndOfBibitem
\bibitem[Richtering and Muller(1995)]{richtering:95}
W.~Richtering and H.~Muller, \emph{Langmuir}, 1995, \textbf{11}, 3699\relax
\mciteBstWouldAddEndPuncttrue
\mciteSetBstMidEndSepPunct{\mcitedefaultmidpunct}
{\mcitedefaultendpunct}{\mcitedefaultseppunct}\relax
\EndOfBibitem
\bibitem[Hunt and Zukoski(1996)]{zukoski:96}
W.~J. Hunt and C.~F. Zukoski, \emph{Langmuir}, 1996, \textbf{12}, 6257\relax
\mciteBstWouldAddEndPuncttrue
\mciteSetBstMidEndSepPunct{\mcitedefaultmidpunct}
{\mcitedefaultendpunct}{\mcitedefaultseppunct}\relax
\EndOfBibitem
\bibitem[Gondret and Petit(1997)]{gondret:97}
P.~Gondret and L.~Petit, \emph{Langmuir}, 1997, \textbf{41}, 1261\relax
\mciteBstWouldAddEndPuncttrue
\mciteSetBstMidEndSepPunct{\mcitedefaultmidpunct}
{\mcitedefaultendpunct}{\mcitedefaultseppunct}\relax
\EndOfBibitem
\bibitem[Greenwood \emph{et~al.}(1997)Greenwood, Luckham, and
  Gregory]{greenwood:97}
R.~Greenwood, P.~F. Luckham and T.~Gregory, \emph{J. Colloid Interface Sci.},
  1997, \textbf{191}, 11\relax
\mciteBstWouldAddEndPuncttrue
\mciteSetBstMidEndSepPunct{\mcitedefaultmidpunct}
{\mcitedefaultendpunct}{\mcitedefaultseppunct}\relax
\EndOfBibitem
\bibitem[Shikata \emph{et~al.}(1998)Shikata, Niwa, and Morishima]{shikata:98}
T.~Shikata, H.~Niwa and Y.~Morishima, \emph{J. Rheol.}, 1998, \textbf{42},
  765\relax
\mciteBstWouldAddEndPuncttrue
\mciteSetBstMidEndSepPunct{\mcitedefaultmidpunct}
{\mcitedefaultendpunct}{\mcitedefaultseppunct}\relax
\EndOfBibitem
\bibitem[Ohtsuki(1983)]{ohtsuki:83}
T.~Ohtsuki, \emph{Physika A}, 1983, \textbf{122}, 212\relax
\mciteBstWouldAddEndPuncttrue
\mciteSetBstMidEndSepPunct{\mcitedefaultmidpunct}
{\mcitedefaultendpunct}{\mcitedefaultseppunct}\relax
\EndOfBibitem
\bibitem[N\"agele and Bergenholtz(1998)]{nagele:98}
G.~N\"agele and J.~Bergenholtz, \emph{J. Chem. Phys.}, 1998, \textbf{108},
  9893\relax
\mciteBstWouldAddEndPuncttrue
\mciteSetBstMidEndSepPunct{\mcitedefaultmidpunct}
{\mcitedefaultendpunct}{\mcitedefaultseppunct}\relax
\EndOfBibitem
\bibitem[Lionberger(2002)]{lionberger:02}
R.~A. Lionberger, \emph{Phys. Rev. E}, 2002, \textbf{65}, 061408\relax
\mciteBstWouldAddEndPuncttrue
\mciteSetBstMidEndSepPunct{\mcitedefaultmidpunct}
{\mcitedefaultendpunct}{\mcitedefaultseppunct}\relax
\EndOfBibitem
\bibitem[Chang and Powell(1993)]{chang:93}
C.~Y. Chang and R.~L. Powell, \emph{J. Fluid. Mech.}, 1993, \textbf{253},
  1\relax
\mciteBstWouldAddEndPuncttrue
\mciteSetBstMidEndSepPunct{\mcitedefaultmidpunct}
{\mcitedefaultendpunct}{\mcitedefaultseppunct}\relax
\EndOfBibitem
\bibitem[Chang and Powell(1994)]{chang:94}
C.~Y. Chang and R.~L. Powell, \emph{J. Rheol.}, 1994, \textbf{38}, 85\relax
\mciteBstWouldAddEndPuncttrue
\mciteSetBstMidEndSepPunct{\mcitedefaultmidpunct}
{\mcitedefaultendpunct}{\mcitedefaultseppunct}\relax
\EndOfBibitem
\bibitem[Chang and Powell(1994)]{powell:94}
C.~Y. Chang and R.~L. Powell, \emph{Phys. Fluids}, 1994, \textbf{6}, 1628\relax
\mciteBstWouldAddEndPuncttrue
\mciteSetBstMidEndSepPunct{\mcitedefaultmidpunct}
{\mcitedefaultendpunct}{\mcitedefaultseppunct}\relax
\EndOfBibitem
\bibitem[Farris(1968)]{farris:68}
R.~J. Farris, \emph{Trans. Soc. Rheol.}, 1968, \textbf{12}, 281\relax
\mciteBstWouldAddEndPuncttrue
\mciteSetBstMidEndSepPunct{\mcitedefaultmidpunct}
{\mcitedefaultendpunct}{\mcitedefaultseppunct}\relax
\EndOfBibitem
\bibitem[Foudazi \emph{et~al.}(2012)Foudazi, Masalova, and
  Malkin]{foudazi:2012}
R.~Foudazi, I.~Masalova and A.~Y. Malkin, \emph{J. Rheol.}, 2012, \textbf{56},
  1299\relax
\mciteBstWouldAddEndPuncttrue
\mciteSetBstMidEndSepPunct{\mcitedefaultmidpunct}
{\mcitedefaultendpunct}{\mcitedefaultseppunct}\relax
\EndOfBibitem
\bibitem[Sentjabrskaja \emph{et~al.}()Sentjabrskaja, Laurati, Egelhaaf, and
  {Th. Voigtmann}]{sentjabrskaya_linear_2012}
T.~Sentjabrskaja, M.~Laurati, S.~U. Egelhaaf and {Th. Voigtmann}, \emph{in
  preparation.}\relax
\mciteBstWouldAddEndPunctfalse
\mciteSetBstMidEndSepPunct{\mcitedefaultmidpunct}
{}{\mcitedefaultseppunct}\relax
\EndOfBibitem
\bibitem[Sentjabrskaja \emph{et~al.}(2013)Sentjabrskaja, Egelhaaf, and
  Laurati]{sentjabrskaya_linear_2012_2}
T.~Sentjabrskaja, S.~U. Egelhaaf and M.~Laurati, \emph{AIP Conf. Proc.}, 2013,
  \textbf{1518}, 206\relax
\mciteBstWouldAddEndPuncttrue
\mciteSetBstMidEndSepPunct{\mcitedefaultmidpunct}
{\mcitedefaultendpunct}{\mcitedefaultseppunct}\relax
\EndOfBibitem
\bibitem[Pham \emph{et~al.}(2006)Pham, Petekidis, Vlassopoulos, Egelhaaf,
  Pusey, and Poon]{pham06}
K.~N. Pham, G.~Petekidis, D.~Vlassopoulos, S.~U. Egelhaaf, P.~N. Pusey and
  W.~C.~K. Poon, \emph{Europhys. Lett.}, 2006, \textbf{75}, 624--630\relax
\mciteBstWouldAddEndPuncttrue
\mciteSetBstMidEndSepPunct{\mcitedefaultmidpunct}
{\mcitedefaultendpunct}{\mcitedefaultseppunct}\relax
\EndOfBibitem
\bibitem[Pham \emph{et~al.}(2008)Pham, Petekidis, Vlassopoulos, Egelhaaf, Poon,
  and Pusey]{pham08}
K.~N. Pham, G.~Petekidis, D.~Vlassopoulos, S.~U. Egelhaaf, W.~C.~K. Poon and
  P.~N. Pusey, \emph{J. Rheol.}, 2008, \textbf{52}, 649--676\relax
\mciteBstWouldAddEndPuncttrue
\mciteSetBstMidEndSepPunct{\mcitedefaultmidpunct}
{\mcitedefaultendpunct}{\mcitedefaultseppunct}\relax
\EndOfBibitem
\bibitem[Laurati \emph{et~al.}(2011)Laurati, Egelhaaf, and
  Petekidis]{laurati:jor:2011}
M.~Laurati, S.~U. Egelhaaf and G.~Petekidis, \emph{J. Rheol.}, 2011,
  \textbf{55}, 673--706\relax
\mciteBstWouldAddEndPuncttrue
\mciteSetBstMidEndSepPunct{\mcitedefaultmidpunct}
{\mcitedefaultendpunct}{\mcitedefaultseppunct}\relax
\EndOfBibitem
\bibitem[Koumakis and Petekidis(2011)]{koumakis:soft:2011}
N.~Koumakis and G.~Petekidis, \emph{Soft Matter}, 2011, \textbf{7},
  2456--2470\relax
\mciteBstWouldAddEndPuncttrue
\mciteSetBstMidEndSepPunct{\mcitedefaultmidpunct}
{\mcitedefaultendpunct}{\mcitedefaultseppunct}\relax
\EndOfBibitem
\bibitem[Petekidis \emph{et~al.}(2002)Petekidis, Moussaid, and
  Pusey]{petekidis02}
G.~Petekidis, A.~Moussaid and P.~N. Pusey, \emph{Phys. Rev. E}, 2002,
  \textbf{66}, 051402\relax
\mciteBstWouldAddEndPuncttrue
\mciteSetBstMidEndSepPunct{\mcitedefaultmidpunct}
{\mcitedefaultendpunct}{\mcitedefaultseppunct}\relax
\EndOfBibitem
\bibitem[Petekidis \emph{et~al.}(2003)Petekidis, Vlassopoulos, and
  Pusey]{petekidis03}
G.~Petekidis, D.~Vlassopoulos and P.~N. Pusey, \emph{Faraday Discuss}, 2003,
  \textbf{123}, 287\relax
\mciteBstWouldAddEndPuncttrue
\mciteSetBstMidEndSepPunct{\mcitedefaultmidpunct}
{\mcitedefaultendpunct}{\mcitedefaultseppunct}\relax
\EndOfBibitem
\bibitem[Yethiraj and {van Blaaderen}(2003)]{yethiraj03}
A.~Yethiraj and A.~{van Blaaderen}, \emph{Nature}, 2003, \textbf{421},
  513--517\relax
\mciteBstWouldAddEndPuncttrue
\mciteSetBstMidEndSepPunct{\mcitedefaultmidpunct}
{\mcitedefaultendpunct}{\mcitedefaultseppunct}\relax
\EndOfBibitem
\bibitem[Jenkins and Egelhaaf(2008)]{jenkins08}
M.~C. Jenkins and S.~U. Egelhaaf, \emph{Adv. \ Coll. \ Interface \ Sci.}, 2008,
  \textbf{136}, 65--92\relax
\mciteBstWouldAddEndPuncttrue
\mciteSetBstMidEndSepPunct{\mcitedefaultmidpunct}
{\mcitedefaultendpunct}{\mcitedefaultseppunct}\relax
\EndOfBibitem
\bibitem[Schaertl and Silescu(1994)]{schaertl94}
W.~Schaertl and H.~Silescu, \emph{J. \ Stat. \ Phys.}, 1994, \textbf{77},
  1007--1025\relax
\mciteBstWouldAddEndPuncttrue
\mciteSetBstMidEndSepPunct{\mcitedefaultmidpunct}
{\mcitedefaultendpunct}{\mcitedefaultseppunct}\relax
\EndOfBibitem
\bibitem[Poon \emph{et~al.}(2012)Poon, Weeks, and Royall]{Poon/Weeks/Royall}
W.~C.~K. Poon, E.~R. Weeks and C.~P. Royall, \emph{Soft Matter}, 2012,
  \textbf{8}, 21--30\relax
\mciteBstWouldAddEndPuncttrue
\mciteSetBstMidEndSepPunct{\mcitedefaultmidpunct}
{\mcitedefaultendpunct}{\mcitedefaultseppunct}\relax
\EndOfBibitem
\bibitem[Koumakis \emph{et~al.}(2012)Koumakis, Pamvouxoglou, Poulos, and
  Petekidis]{koumakis:2012}
N.~Koumakis, A.~Pamvouxoglou, A.~S. Poulos and G.~Petekidis, \emph{Soft
  Matter}, 2012, \textbf{8}, 4271--4284\relax
\mciteBstWouldAddEndPuncttrue
\mciteSetBstMidEndSepPunct{\mcitedefaultmidpunct}
{\mcitedefaultendpunct}{\mcitedefaultseppunct}\relax
\EndOfBibitem
\bibitem[Sch{\"o}pe \emph{et~al.}(2007)Sch{\"o}pe, Bryant, and van
  Megen]{schoepe:2007}
H.~J. Sch{\"o}pe, G.~Bryant and W.~van Megen, \emph{J. Chem. Phys.}, 2007,
  \textbf{127}, 084505\relax
\mciteBstWouldAddEndPuncttrue
\mciteSetBstMidEndSepPunct{\mcitedefaultmidpunct}
{\mcitedefaultendpunct}{\mcitedefaultseppunct}\relax
\EndOfBibitem
\bibitem[Koumakis \emph{et~al.}(2012)Koumakis, Laurati, Egelhaaf, Brady, and
  Petekidis]{nick:prl:12}
N.~Koumakis, M.~Laurati, S.~U. Egelhaaf, J.~F. Brady and G.~Petekidis,
  \emph{Phys. Rev. Lett.}, 2012, \textbf{108}, 098303\relax
\mciteBstWouldAddEndPuncttrue
\mciteSetBstMidEndSepPunct{\mcitedefaultmidpunct}
{\mcitedefaultendpunct}{\mcitedefaultseppunct}\relax
\EndOfBibitem
\bibitem[Smith \emph{et~al.}(2007)Smith, Petekidis, Egelhaaf, and
  Poon]{smith:pre:2007}
P.~A. Smith, G.~Petekidis, S.~U. Egelhaaf and W.~C.~K. Poon, \emph{Phys. Rev.
  E}, 2007, \textbf{76}, 041402\relax
\mciteBstWouldAddEndPuncttrue
\mciteSetBstMidEndSepPunct{\mcitedefaultmidpunct}
{\mcitedefaultendpunct}{\mcitedefaultseppunct}\relax
\EndOfBibitem
\bibitem[Conrad \emph{et~al.}(2010)Conrad, Wyss, Trappe, Manley, Miyazaki,
  Kaufman, Schofield, Reichman, and Weitz]{conrad:421}
J.~C. Conrad, H.~M. Wyss, V.~Trappe, S.~Manley, K.~Miyazaki, L.~J. Kaufman,
  A.~B. Schofield, D.~R. Reichman and D.~A. Weitz, \emph{J. Rheol.}, 2010,
  \textbf{54}, 421--438\relax
\mciteBstWouldAddEndPuncttrue
\mciteSetBstMidEndSepPunct{\mcitedefaultmidpunct}
{\mcitedefaultendpunct}{\mcitedefaultseppunct}\relax
\EndOfBibitem
\bibitem[K{\"o}nig \emph{et~al.}(2005)K{\"o}nig, Hund, Zahn, and
  Maret]{koenig_epje05}
H.~K{\"o}nig, R.~Hund, K.~Zahn and G.~Maret, \emph{Eur. Phys. J. E}, 2005,
  \textbf{18}, 287--293\relax
\mciteBstWouldAddEndPuncttrue
\mciteSetBstMidEndSepPunct{\mcitedefaultmidpunct}
{\mcitedefaultendpunct}{\mcitedefaultseppunct}\relax
\EndOfBibitem
\bibitem[Hamanaka and Onuki(2007)]{onuki_pre07}
T.~Hamanaka and A.~Onuki, \emph{Phys. Rev. E}, 2007, \textbf{75}, 041503\relax
\mciteBstWouldAddEndPuncttrue
\mciteSetBstMidEndSepPunct{\mcitedefaultmidpunct}
{\mcitedefaultendpunct}{\mcitedefaultseppunct}\relax
\EndOfBibitem
\bibitem[Yunker \emph{et~al.}(2010)Yunker, Zhang, and Yodh]{yunker_prl10}
P.~Yunker, Z.~Zhang and A.~G. Yodh, \emph{Phys. Rev. Lett.}, 2010,
  \textbf{104}, 015701\relax
\mciteBstWouldAddEndPuncttrue
\mciteSetBstMidEndSepPunct{\mcitedefaultmidpunct}
{\mcitedefaultendpunct}{\mcitedefaultseppunct}\relax
\EndOfBibitem
\bibitem[Ebert \emph{et~al.}(2008)Ebert, Keim, and Maret]{ebert:2008}
F.~Ebert, P.~Keim and G.~Maret, \emph{Eur. Phys. J. E}, 2008, \textbf{26},
  161\relax
\mciteBstWouldAddEndPuncttrue
\mciteSetBstMidEndSepPunct{\mcitedefaultmidpunct}
{\mcitedefaultendpunct}{\mcitedefaultseppunct}\relax
\EndOfBibitem
\bibitem[Bonales \emph{et~al.}(2012)Bonales, Martinez-Pedrero, Rubio, Rubio,
  and Ortega]{bonales:2012}
L.~J. Bonales, F.~Martinez-Pedrero, M.~A. Rubio, R.~G. Rubio and F.~Ortega,
  \emph{Langmuir}, 2012, \textbf{28}, 16555\relax
\mciteBstWouldAddEndPuncttrue
\mciteSetBstMidEndSepPunct{\mcitedefaultmidpunct}
{\mcitedefaultendpunct}{\mcitedefaultseppunct}\relax
\EndOfBibitem
\bibitem[Biazzo \emph{et~al.}(2009)Biazzo, Caltagirone, Parisi, and
  Zamponi]{biazzo:2009}
I.~Biazzo, F.~Caltagirone, G.~Parisi and F.~Zamponi, \emph{Phys. Rev. Lett.},
  2009, \textbf{102}, 195701\relax
\mciteBstWouldAddEndPuncttrue
\mciteSetBstMidEndSepPunct{\mcitedefaultmidpunct}
{\mcitedefaultendpunct}{\mcitedefaultseppunct}\relax
\EndOfBibitem
\bibitem[Farr and Groot(2009)]{farr:2009}
R.~S. Farr and R.~D. Groot, \emph{J. Chem. Phys.}, 2009, \textbf{131},
  244104\relax
\mciteBstWouldAddEndPuncttrue
\mciteSetBstMidEndSepPunct{\mcitedefaultmidpunct}
{\mcitedefaultendpunct}{\mcitedefaultseppunct}\relax
\EndOfBibitem
\bibitem[Koumakis \emph{et~al.}(2008)Koumakis, Schofield, and
  Petekidis]{koumakis:2008}
N.~Koumakis, A.~B. Schofield and G.~Petekidis, \emph{Soft Matter}, 2008,
  \textbf{4}, 2008--2018\relax
\mciteBstWouldAddEndPuncttrue
\mciteSetBstMidEndSepPunct{\mcitedefaultmidpunct}
{\mcitedefaultendpunct}{\mcitedefaultseppunct}\relax
\EndOfBibitem
\bibitem[Mayer \emph{et~al.}(2008)Mayer, Zaccarelli, Stiakakis, Likos,
  Sciortino, Munam, Gauthier, Hadjichristidis, Iatrou, Tartaglia, L{\"o}wen,
  and Vlassopoulos]{mayer:2008}
C.~Mayer, E.~Zaccarelli, E.~Stiakakis, C.~N. Likos, F.~Sciortino, A.~Munam,
  M.~Gauthier, N.~Hadjichristidis, H.~Iatrou, P.~Tartaglia, H.~L{\"o}wen and
  D.~Vlassopoulos, \emph{Nat. Mater.}, 2008, \textbf{7}, 780\relax
\mciteBstWouldAddEndPuncttrue
\mciteSetBstMidEndSepPunct{\mcitedefaultmidpunct}
{\mcitedefaultendpunct}{\mcitedefaultseppunct}\relax
\EndOfBibitem
\end{mcitethebibliography}
\bibliographystyle{rsc} 
}

\end{document}